\documentclass{acmtog}

\usepackage{graphicx}
\usepackage{amsmath}
\usepackage{amssymb}
\usepackage{color}
\usepackage[abs]{overpic}
\usepackage{rotating}
\usepackage{colortbl}
\usepackage{xcolor}

\definecolor{darkyellow}{rgb}{1,.9,0}

\def\argmin{\mathop{\rm argmin}}

\newcommand{\boldhead}[1]{\vspace{0.05in}\noindent\textbf{#1.}}

\newcommand{\ve}[1]{{\bf #1}}
\newcommand{\comment}[1]{}

\acmVolume{VV}
\acmNumber{N}
\acmYear{YYYY}
\acmMonth{Month}
\acmArticleNum{XXX}  
\acmdoi{10.1145/XXXXXXX.YYYYYYY}

\acmVolume{28}
\acmNumber{4}
\acmYear{2009}
\acmMonth{August}
\acmArticleNum{106}  
\acmdoi{10.1145/1559755.1559763}

\begin{document}

\markboth{K. Karsch, K. Sunkavalli, S. Hadap, N. Carr, H. Jin, R. Fonte, M. Sittig, D. Forsyth}{Automatic Scene Inference for 3D Object Compositing}

\title{Automatic Scene Inference for 3D Object Compositing}
%\title{Realistic Editing of Legacy Images through Automatic Scene Inference}

\author{Kevin Karsch${}^1$, Kalyan Sunkavalli${}^2$, Sunil Hadap${}^2$, \\ Nathan Carr${}^2$, Hailin Jin${}^2$, Rafael Fonte${}^1$, \\Michael Sittig${}^1$ David Forsyth${}^1$
\affil{${}^1$University of Illinois} \affil{${}^2$Adobe Research}}


%\terms{\todo{Experimentation, Human Factors}}


%\acmformat{Pamplona, V. F., Oliveira, M. M., and Baranoski, G. V. G. 2009. Photorealistic models for pupil light reflex and iridal pattern deformation.  {ACM Trans. Graph.} 28, 4, Article 106 (August 2009), 11 pages.\newline  DOI $=$10.1145/1559755.1559763\newline http://doi.acm.org/10.1145/1559755.1559763}

\maketitle

\begin{bottomstuff} 
%\todo{ Manuel M. Oliveira acknowledges a CNPq-Brazil fellowship (305613/2007-3). Gladimir V. G. Baranoski acknowledges a NSERC-Canada grant (238337). Microsoft Brazil provided additional support.}
Authors' email addresses: \{karsch1,sittig2,dafonte2,daf\}@illinois.edu, \{sunkaval,hadap,ncarr,hljin\}@adobe.com
\end{bottomstuff}

\begin{abstract} 
We present a user-friendly image editing system that supports a drag-and-drop object insertion (where the user merely drags objects into the image, and the system automatically places them in 3D and relights them appropriately), post-process illumination editing, and depth-of-field manipulation. Underlying our system is a fully automatic technique for recovering a comprehensive 3D scene model (geometry, illumination, diffuse albedo and camera parameters) from a single, low dynamic range photograph. This is made possible by two novel contributions: an illumination inference algorithm that recovers a full lighting model of the scene (including light sources that are not directly visible in the photograph), and a depth estimation algorithm that combines data-driven depth transfer with geometric reasoning about the scene layout. A user study shows that our system produces perceptually convincing results, and achieves the same level of realism as techniques that require significant user interaction.
\end{abstract}

%%%%%%%%%%%%%%%%%%%%%%%%%%%%%%%%%%%%%%%%%%%%%%%%
\section{Introduction} 

%----------------------------------------------- Teaser
\begin{figure}[t]
\vspace{-30mm}
\begin{center}
%\begin{overpic}[width=.33\linewidth] {fig/teaser/dragndrop.png}
%\setlength\fboxsep{0pt}
%\setlength\fboxrule{0.5pt}
%\put(-5,60){\fbox{\includegraphics[width=0.17\linewidth]{fig/teaser/input.png}}}
%\end{overpic}
\includegraphics[width=.99\linewidth]{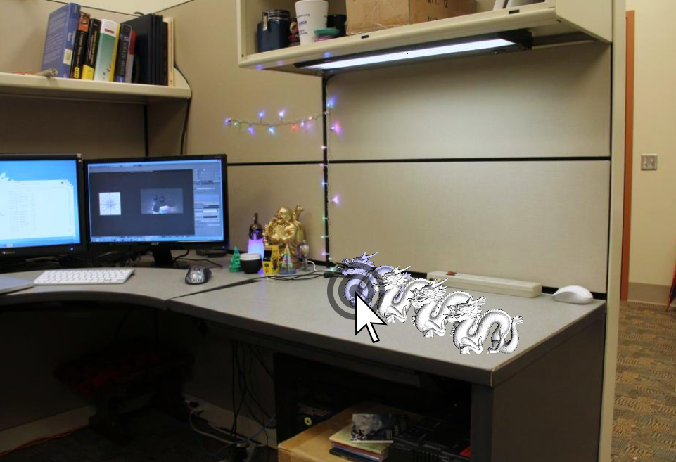}\\ \vspace{0.5mm}
\includegraphics[width=.99\linewidth]{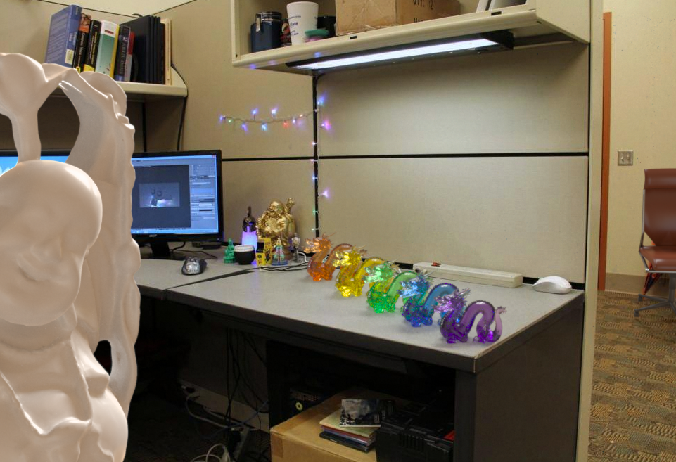}\\ \vspace{0.5mm}
\includegraphics[width=.99\linewidth]{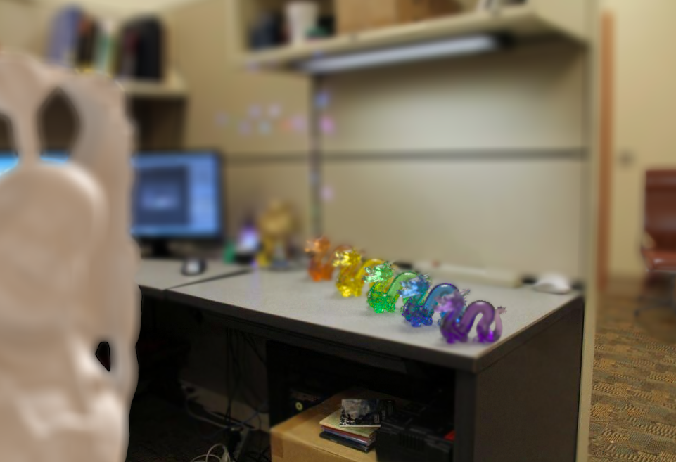}
\end{center}
\vspace{-4mm}
\caption{From a single LDR photograph, our system automatically estimates a 3D scene model without any user interaction or additional information. These scene models facilitate photorealistic, physically grounded image editing operations, which we demonstrate with an intuitive interface. With our system, a user can simply drag-and-drop 3D models into a picture (top), render objects seamlessly into photographs with a single click (middle), adjust the illumination, and refocus the image in real time (bottom). Best viewed in color at high-resolution.}
\label{fig:teaser}
\end{figure}
%----------------------------------------------- Teaser

%----------------------------------------------- Overview
\begin{figure*}
\centerline{\includegraphics[width=0.90\textwidth]{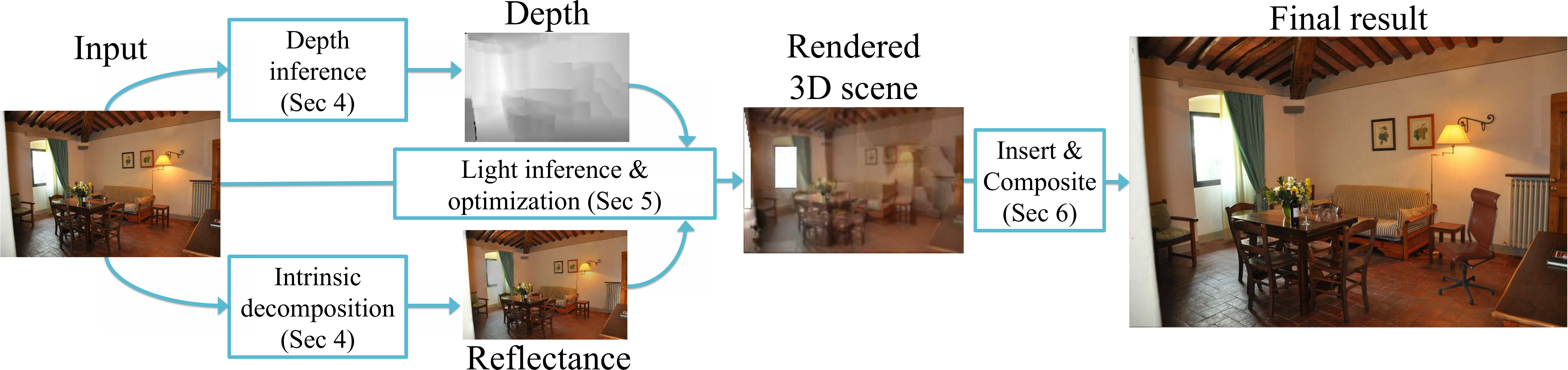}}
\vspace{-2mm}
\caption{Our system allows for physically grounded image editing (e.g., the inserted dragon and chair on the right), facilitated by our automatic scene estimation procedure. To compute a scene from a single image, we automatically estimate dense depth and diffuse reflectance (the geometry and materials of our scene).  Sources of illumination are then inferred without any user input to form a complete 3D scene, conditioned on the estimated scene geometry. Using a simple, drag-and-drop interface, objects are quickly inserted and composited into the input image with realistic lighting, shadowing, and perspective. {\it Photo credits: {\footnotesize \textcopyright{}}Salvadonica Borgo.}
}
\label{fig:overview}
\end{figure*}
%----------------------------------------------- Overview

Many applications require a user to insert 3D characters, props, or other synthetic objects into images. In many existing photo editors, it is the artist's job to create photorealistic effects by recognizing the physical space present in an image. For example, to add a new object into an image, the artist must determine how the object will be lit, where shadows will be cast, and the perspective at which the object will be viewed. In this paper, we demonstrate a new kind of image editor -- one that computes the physical space of the photograph automatically, allowing an artist (or, in fact, anyone) to make physically grounded edits with only a few mouse clicks.

Our system works by inferring the physical scene (geometry, illumination, etc.) that corresponds to a single LDR photograph. This process is fully automatic, requires no special hardware, and works for legacy images.  We show that our inferred scene models can be used to facilitate a variety of physically-based image editing operations. For example, objects can be seamlessly inserted into the photograph, light source intensity can be modified, and the picture can be refocused on the fly. Achieving these edits with existing software is a painstaking process that takes a great deal of artistry and expertise. 

In order to facilitate realistic object insertion and rendering we need to hypothesize camera parameters, scene geometry, surface materials, and sources of illumination. To address this, we develop a new method for both single-image depth and illumination inference. We are able to build a full 3D scene model without any user interaction, including camera parameters and reflectance estimates. 

%We describe how to estimate scene geometry and materials in Sec~\ref{sec:recon}, followed by detailing a technique for estimating illumination in Sec~\ref{sec:illum}. Our system provides an interface for editing images using our automatically generated scene models (Sec~\ref{sec:interface}).  We have evaluated our results with a user study (Sec~\ref{sec:study}). Figure~\ref{fig:overview} illustrates the pipeline of our system. 

\boldhead{Contributions}
Our primary contribution is a completely automatic algorithm for estimating a full 3D scene model from a single LDR photograph. Our system contains two technical contributions: illumination inference and depth estimation. We have developed a novel, data-driven illumination estimation procedure that automatically estimates a physical lighting model for the entire scene (including out-of-view light sources). This estimation is aided by our single-image light classifier to detect emitting pixels, which we believe is the first of its kind. We also demonstrate state-of-the-art depth estimates by combining data-driven depth inference with geometric reasoning.

We have created an intuitive interface for inserting 3D models seamlessly into photographs, using our scene approximation method to relight the object and facilitate drag-and-drop insertion. Our interface also supports other physically grounded image editing operations, such as post-process depth-of-field and lighting changes. In a user study, we show that our system is capable of making photorealistic edits: in side-by-side comparisons of ground truth photos with photos edited by our software, subjects had a difficult time choosing the ground truth.

\boldhead{Limitations} 
Our method works best when scene lighting is diffuse, and therefore generally works better indoors than out (see our user studies and results in Sec~\ref{sec:study}). Our scene models are clearly not canonical representations of the imaged scene and often differ significantly from the true scene components. These coarse scene reconstructions suffice in many cases to produce realistically edited images. However, in some case, errors in either geometry, illumination, or materials may be stark enough to manifest themselves in unappealing ways while editing. For example, inaccurate geometry could cause odd looking shadows for inserted objects, and inserting light sources can exacerbate geometric errors. 
%Geometric errors can typically be avoided by placing objects in specific locations (where the geometry is more accurate), but this is clearly not desirable if an object must be placed in a certain location in the image.
%However, we have found it reasonably easy to avoid these issues, and given the drag-and-drop nature of our application, generating new/different edits is extremely fast.
Also, our editing software does not handle object insertion {\it behind} existing scene elements automatically, and cannot be used to deblur an image taken with wide aperture. A Manhattan World is assumed in our camera pose and depth estimation stages, but our method is still applicable in scenes where this assumption does not hold (see Fig~\ref{fig:results}).

%%%%%%%%%%%%%%%%%%%%%%%%%%%%%%%%%%%%%%%%%%%%%%%%
\section{Related work}
In order to build a physically based image editor (one that supports operations such as lighting-consistent object insertion, relighting, and new view synthesis), it is requisite to model the three major factors in image formation: geometry, illumination, and surface reflectance. Existing approaches are prohibitive to most users as they require either manually recreating or measuring an imaged scene with hardware aids~\cite{Debevec:98,Yu:1999,Boivin:2001,Debevec05:Parthenon}.
%Images are formed as the result of illumination interacting with scene geometry and material reflectances in complex ways. Explicitly measuring (or alternatively, modeling) each of these factors can enable physically-based operations such as lighting-consistent object insertion, relighting, and new view synthesis~\cite{Debevec:98,Yu:1999,Debevec05:Parthenon}, but acquisition setups required for such approaches make them prohibitive for most users. 
In contrast, Lalonde et al.~\shortcite{Lalonde:2007uj} and Karsch et al.~\shortcite{Karsch:SA2011} have shown that even coarse estimates of scene geometry, reflectance properties, illumination, and camera parameters are sufficient for many image editing tasks. Their techniques require a user to model the scene geometry and illumination -- a task that requires time and an understanding of 3D authoring tools. While our work is similar in spirit to theirs, our technique is fully automatic and is still able to produce results with the same perceptual quality.

Similar to our method, Barron and Malik~\shortcite{Barron2013A} recover shape, surface albedo and illumination for entire scenes, but their method requires a coarse input depth map (e.g. from a Kinect) and is not directly suitable for object insertion as illumination is only estimated near surfaces (rather than the entire volume).

\boldhead{Geometry} User-guided approaches to single image modeling~\cite{Horry97:TiP,Criminisi00:SVMetrology,Oh01:IBM} have been successfully used to create 3D reconstructions that allow for viewpoint variation. Single image depth estimation techniques have used learned relationships between image features and geometry to estimate depth~\cite{Hoiem05:Popup,Saxena:pami09,Liu:cvpr10,Karsch:ECCV12}. Our depth estimation technique improves upon these methods by incorporating geometric constraints, using intuition from past approaches which estimate depth by assuming a Manhattan World (\cite{Delage:ISRR:05} for single images, and ~\cite{Furukawa:09,Gallup:CVPR:10} for multiple images).

Other approaches to image modeling have explicitly parametrized indoor scenes as 3D boxes (or as collections of orthogonal planes)~\cite{Lee:CVPR09,Hedau:2009vi,Karsch:SA2011,Schwing:ECCV12}. In our work, we use appearance features to infer image depth, but augment this inference with priors based on geometric reasoning about the scene. 

The contemporaneous works of Satkin et al.~\cite{satkin_bmvc2012} and Del Pero et al.~\cite{pero:cvpr:13} predict 3D scene reconstructions for the rooms and their furniture. Their predicted models can be very good semantically, but are not suited well for our editing applications as the models are typically not well-aligned with the edges and boundaries in the image.

%While all these techniques seek to estimate image depth, shape-from-shading algorithms (SFS)~\cite{Horn70:SfS,Zhang99:SfS,Johnson:CVPR:2011} use the relationship between illumination and scene shading to recover geometry in the form of surface normals. In our work, we first estimate the scene geometry and then use it to infer the illumination in the image from the shading.

%Previously, there has been much work on estimating low-order representations of indoor scenes~\cite{Lee:CVPR09,Hedau:2009vi,Schwing:ECCV12}. These methods assume a Manhattan World (i.e. that most planes lie in one of three mutually orthogonal orientations), which tends to be a reasonable assumption for indoor scenes~\cite{manhattan}, and attempt to explain the imaged scene's geometry with a simple 3D cuboid. A similar representation has been used for object insertion in the semiautomatic relighting work of Karsch et al.~\shortcite{Karsch:SA2011}. 

%Single image depth estimation has also seen much progress lately~\cite{Saxena:pami09,Liu:cvpr10,Karsch:ECCV12}, but existing methods are purely example-driven, and incorporate little to no geometric information during training or inference. We combine insights from these past techniques, and develop a dense depth estimation algorithm that is guided by geometric cues extracted automatically from the input image. The resulting depth tends to be piecewise planar, and can be significantly more accurate than past approaches.
	
\boldhead{Illumination} Lighting estimation algorithms vary by the representation they use for illumination. Point light sources in the scene can be detected by analyzing silhouettes and shading along object contours~\cite{Johnson05:Lighting,LopezMoreno2010698}. Lalonde et al.~\shortcite{Lalonde:2009vp} use a physically-based model for sky illumination and a data-driven sunlight model to recover an environment map from time-lapse sequences. In subsequent work, they use the appearance of the sky in conjunction with cues such as shadows and shading to recover an environment map from a single image~\cite{Lalonde:2009vp}. Nishino et al.~\shortcite{Nishino04:EyesRelighting} recreate environment maps of the scene from reflections in eyes. Johnson and Farid~\shortcite{Johnson07:Complex} estimate lower-dimensional spherical harmonics-based lighting models from images. Panagopoulos et al.~\shortcite{Panagopoulos:cvpr:11} show that automatically detected shadows can be used to recover an illumination environment from a single image, but require coarse geometry as input.

While all these techniques estimate physically-based lighting from the scene, Khan et al.~\shortcite{Khan:2006:IME:1179352.1141937} show that wrapping an image to create the environment map can suffice for certain applications. %, and we compare our illumination to this method and that of Karsch et al.~\shortcite{Karsch:SA2011} in Fig~\ref{fig:comparison}. 

Our illumination estimation technique attempts to predict illumination both within and outside the photograph's frustum with a data-driven matching approach; such approaches have seen previous success in recognizing scene viewpoint~\cite{Xiao:CVPR:12} and view extrapolation~\cite{Zhang:CVPR:13}.

We also attempt to predict a one-parameter camera response function jointly during our inverse rendering optimization. Other processes exist for recovering camera response, but require multiple of images~\cite{Diaz:2013wk}.

\boldhead{Materials} In order to infer illumination, we separate the input image into diffuse albedo and shading using the Color Retinex algorithm~\cite{Grosse:iccv:09}. We assume that the scene is Lambertian, and find that this suffices for our applications. Other researchers have looked at the problem of recovering the Bi-directional Reflectance Density Function (BRDF) from a single image, but as far as we know, there are no such methods that work automatically and at the scene (as opposed to object) level. These techniques typically make the problem tractable by using low-dimensional representations for the BRDF such as spherical harmonics~\cite{Ramamoorthi04:SignalProcessing}, bi-variate models for isotropic reflectances~\cite{Romeiro08:PR,Romeiro10:BR}, and data-driven statistical models~\cite{Lombardi12:ReflectanceIllumination,Lombardi12:MultiMaterial}. All these techniques require the shape to be known and in addition, either require the illumination to be given, or use priors on lighting to constrain the problem space. 

%Barron and Malik~\shortcite{Barron12:SIAlbedo,Barron12:SIReflectance} simultaneously estimate shape, surface albedo and illumination from a single image. This is a highly under-constrained problem, and they incorporate priors on albedo, shape, and illumination to make the solution converge to the most likely set of parameters.

\boldhead{Perception} Even though our estimates of scene geometry, materials, and illumination are coarse, they enable us to create realistic composites. This is possible because even large changes in lighting are often not perceivable to the human visual system. This has been shown to be true for both point light sources~\cite{LopezMoreno2010698} and complex illumination~\cite{Ramanarayanan07:VisualEq}.

%%%%%%%%%%%%%%%%%%%%%%%%%%%%%%%%%%%%%%%%%%%%%%%%
\section{Method overview}
\label{sec:overview}
Our method consists of three primary steps, outlined in Fig~\ref{fig:overview}. First, we estimate the physical space of the scene (camera parameters and geometry), as well as the per-pixel diffuse reflectance (Sec~\ref{sec:recon}). Next, we estimate scene illumination (Sec~\ref{sec:illum}) which is guided by our previous estimates (camera, geometry, reflectance). Finally, our interface is used to composite objects, improve illumination estimates, or change the depth-of-field (Sec~\ref{sec:interface}). We have evaluated our method with a large-scale user study (Sec~\ref{sec:study}), and additional details and results can be found in the corresponding supplemental document. Figure~\ref{fig:overview} illustrates the pipeline of our system. 

\boldhead{Scene parameterization} Our geometry is in the form of a depth map, which is triangulated to form a polygonal mesh (depth is unprojected according to our estimated pinhole camera). Our illumination model contains polygonal area sources, as well as one or more spherical image-based lights.

While unconventional, our models are suitable for most off-the-shelf rendering software, and we have found our models to produce better looking estimates than simpler models (e.g. planar geometry with infinitely distant lighting).

\boldhead{Automatic indoor/outdoor scene classification} As a pre-processing step, we automatically detect whether the input image is indoors or outdoors. We use a simple method: $k$-nearest-neighbor matching of GIST features~\cite{Oliva:2001wt} between the input image and all images from the indoor NYUv2 dataset and the outdoor Make3D Dataset. We choose $k=7$, and decide use majority-voting to determine if the image is indoors or outdoors (e.g. if 4 of the nearest neighbors are from the Make3D dataset, we consider it to be outdoors). More sophisticated methods could also work.

Our method uses different training images and classifiers depending on whether the input image is classified as an indoor or outdoor scene. 

%%%%%%%%%%%%%%%%%%%%%%%%%%%%%%%%%%%%%%%%%%%%%%%%
\section{Single image reconstruction}
\label{sec:recon}

%----------------------------------------------- Depth overview
\begin{figure}
\centerline{\includegraphics[width=0.95\linewidth]{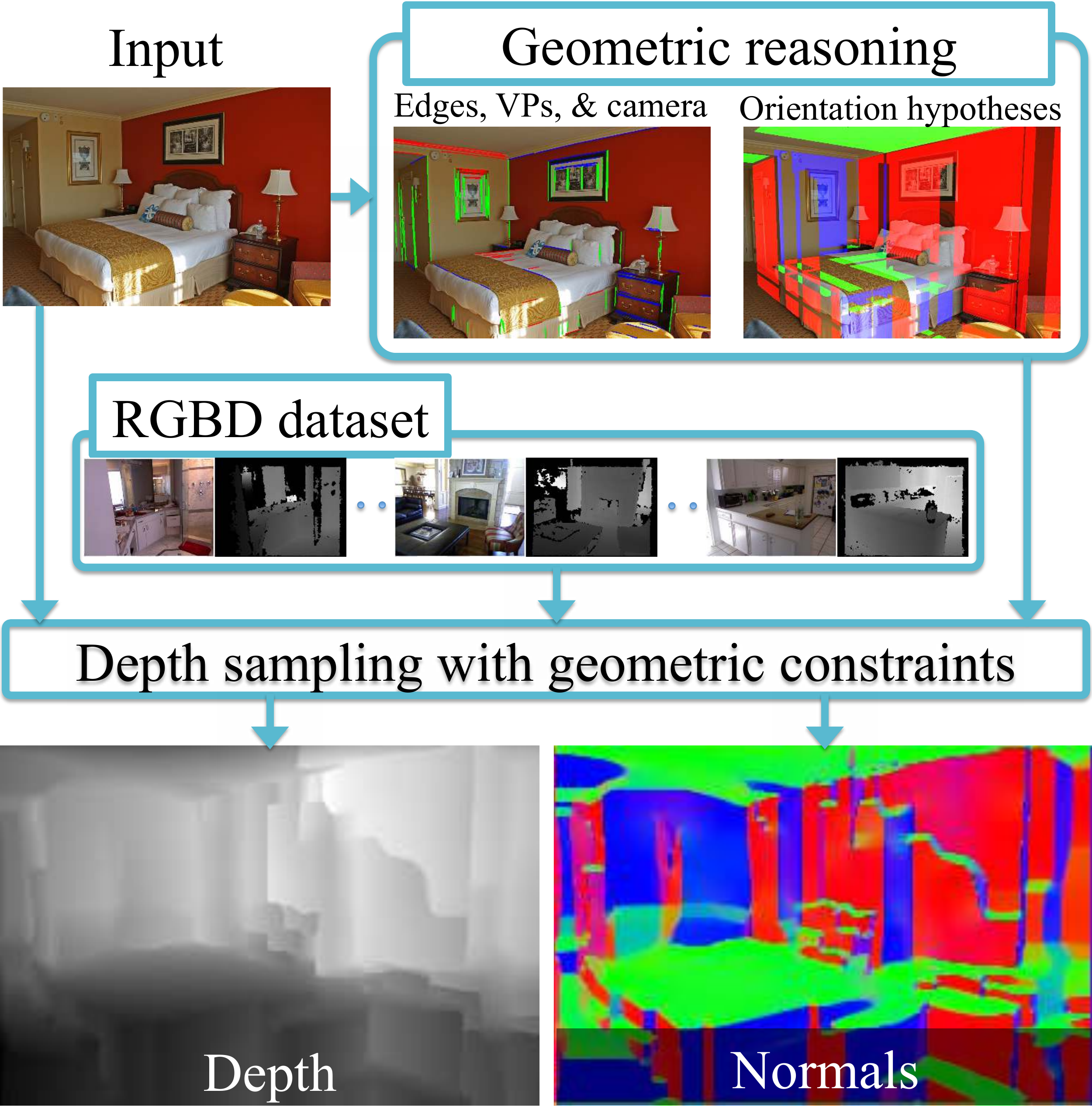}}
\caption{Automatic depth estimation algorithm. Using the geometric reasoning method of Lee et al.~\protect\shortcite{Lee:CVPR09}, we estimate focal length and a sparse surface orientation map. Facilitated by a dataset of RGBD images, we then apply a non-parametric depth sampling approach to compute the per-pixel depth of the scene. The geometric cues are used during inference to enforce orientation constraints, piecewise-planarity, and surface smoothness. The result is a dense reconstruction of the scene that is suitable for realistic, physically grounded editing. {\it Photo credits: Flickr user {\footnotesize \textcopyright{}}``Mr.TinDC''.}
}
\label{fig:depth_overview}
\end{figure}
%----------------------------------------------- Depth overview

The first step in our algorithm is to estimate the physical space of the scene, which we encode with a depth map, camera parameters, and spatially-varying diffuse materials.  Here, we describe how to estimate these components, including a new technique for estimating depth from a single image that adheres to geometric intuition about indoor scenes.

\boldhead{Single image depth estimation} 
Karsch et al.~\shortcite{Karsch:ECCV12} describe a non-parametric, ``depth transfer'' approach for estimating dense, per-pixel depth from a single image. While shown to be state-of-the-art, this method is purely data-driven, and incorporates no explicit geometric information present in many photographs. It requires a database of RGBD (RGB+depth) images, and transfers depth from the dataset to a novel input image in a non-parametric fashion using correspondences in appearance. This method has been shown to work well both indoors and outdoors; however, only appearance cues are used (multi-scale SIFT features), and we have good reason to believe that adding geometric information will aid in this task.

A continuous optimization problem is solved to find the most likely estimate of depth given an input image. In summary, images in the RGBD database are matched to the input and warped so that SIFT features are aligned, and an objective function is minimized to arrive at a solution. We denote $\ve{D}$ as the depth map we wish to infer, and following the notation of Karsch et al., we write the full objective here for completeness:
\begin{equation}
\argmin_{\ve{D}} E(\ve{D}) = \sum_{i\in \text{pixels}} E_t(\ve{D}_i) + \alpha E_s(\ve{D}_i) + \beta E_p(\ve{D}_i),
\end{equation}
where $E_t$ is the data term (depth transfer), $E_s$ enforces smoothness, and $E_p$ is a prior encouraging depth to look like the average depth in the dataset. We refer the reader to~\cite{Karsch:ECCV12} for details.

Our idea is to reformulate the depth transfer objective function and infuse it with geometric information extracted using the geometric reasoning algorithm of Lee et al.~\shortcite{Lee:CVPR09}. Lee et al. detect vanishing points and lines from a single image, and use these to hypothesize a set of sparse surface orientations for the image. The predicted surface orientations are aligned with one of the three dominant directions in the scene (assuming a Manhattan World).

%\boldhead{Geometric reasoning} We use the geometric reasoning algorithm of Lee et al.~\shortcite{Lee:CVPR09} to estimate three dominant, mutually orthogonal scene directions, as well a sparse set of predicted surface orientations (along one of these three directions). In summary, this algorithm detects long, straight lines in an image and classifies them into three clusters (excluding outliers) using a variant of RANSAC. Each cluster of line segments intersect at a vanishing point, which are assumed to be mutually orthogonal. That is, line segments within the same group will will be parallel in 3D, and line segments in different groups are orthogonal. Using these directions and groups of lines as guides, a sparse set of planes oriented along these directions are generated. Fig.~\ref{fig:geometric_reason} shows an example of the estimates lines and planes from an image.

%----------------------------------------------- Depth comparison
\begin{figure}
\centerline{\includegraphics[width=.95\linewidth]{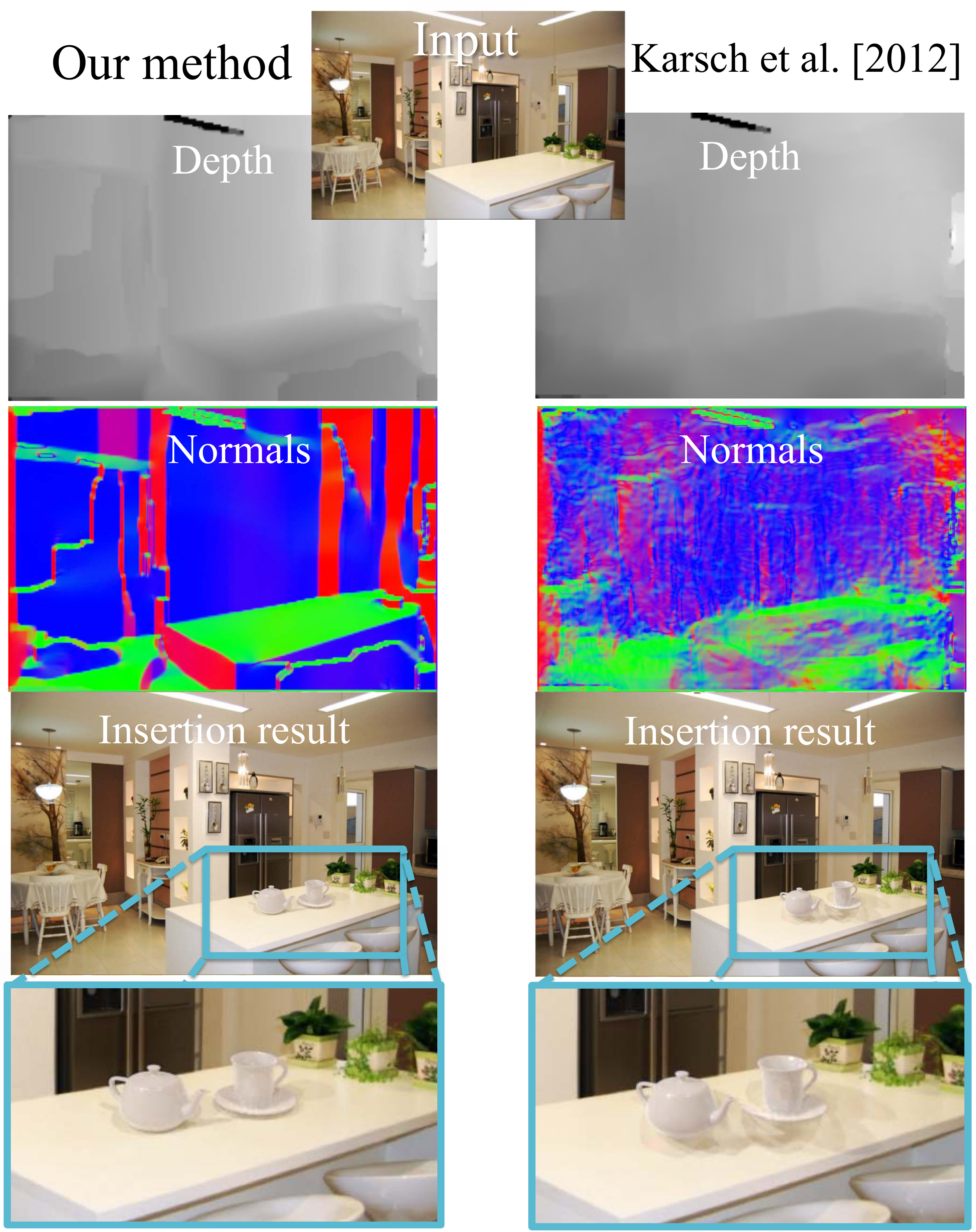}}
\caption{Comparison of different depth estimation techniques. Although the depth maps of our technique and of Karsch et al.~\protect\shortcite{Karsch:ECCV12} appear roughly similar, the surface orientations provide a sense of how distinct the methods are. Our depth estimation procedure (aided by geometric reasoning) is crucial in achieving realistic insertion results, as the noisy surface orientations from Karsch et al. can cause implausible cast shadows and lighting effects.
%(evidenced especially by the cast shadows of inserted objects), and demonstrates the importance of incorporating geometric cues in depth estimation.
}
\label{fig:depth_comparison}
\end{figure}
%----------------------------------------------- Depth comparison

We remove the image-based smoothness ($E_s$) and prior terms ($E_p$), and replace them with geometric-based priors. We add terms to enforce a Manhattan World ($E_m$), constrain the orientation of planar surfaces ($E_o$), and impose 3D smoothness ($E_{3s}$, spatial smoothness in 3D rather than 2D):
\begin{align}
\label{eq:depth}
\argmin_{\ve{D}} E_{geom}(\ve{D}) = \sum_{i\in \text{pixels}} &E_t(\ve{D}_i) + \lambda_m E_m(N(\ve{D})) \ + \\
&\lambda_o E_o(N(\ve{D})) + \lambda_{3s} E_{3s}(N(\ve{D})), \nonumber
\end{align}
where the weights are trained using a coarse-to-fine grid search on held-out ground truth data (indoors: $\lambda_m = 100$, $\lambda_o = 0.5$, $\lambda_{3s} = 0.1$, outdoors: $\lambda_m = 200$, $\lambda_o = 10$, $\lambda_{3s} = 1$); these weights dictate the amount of influence each corresponding prior has during optimization. Descriptions of these priors and implementation details can be found in the supplemental file.

Figure~\ref{fig:depth_overview} shows the pipeline of our depth estimation algorithm, and Fig~\ref{fig:depth_comparison} illustrates the differences between our method and the depth transfer approach; in particular, noisy surface orientations from the depth of Karsch et al.~\shortcite{Karsch:ECCV12} lead to unrealistic insertion and relighting relights.

In supplemental material, we show additional results, including state-of-the-art results on two benchmark datasets using our new depth estimator. 

\boldhead{Camera parameters}
It is well known how to compute a simple pinhole camera (focal length, $f$ and camera center, $(c_0^x, c_0^y)$) and extrinsic parameters from three orthogonal vanishing points~\cite{Hartley:2003te} (computed during depth estimation), and we use this camera model at render-time. 
%\begin{equation} 
%K = \left[ \begin{array}{c c c} f & 0 & c_0^x \\ 0 & f & c_0^y \\ 0 & 0 & 1 \end{array}  \right] \nonumber  
%\end{equation} 

\boldhead{Surface materials} We use Color Retinex (as described in~\cite{Grosse:iccv:09}), to estimate a spatially-varying diffuse material albedo for each pixel in the visible scene. 
%We have found even this simple model of material is good enough for many cases of synthetic object insertion, and 

%%%%%%%%%%%%%%%%%%%%%%%%%%%%%%%%%%%%%%%%%%%%%%%%
\section{Estimating Illumination}
\label{sec:illum}

%----------------------------------------------- Light overview
\begin{figure*}
\centerline{\includegraphics[width=.95\linewidth]{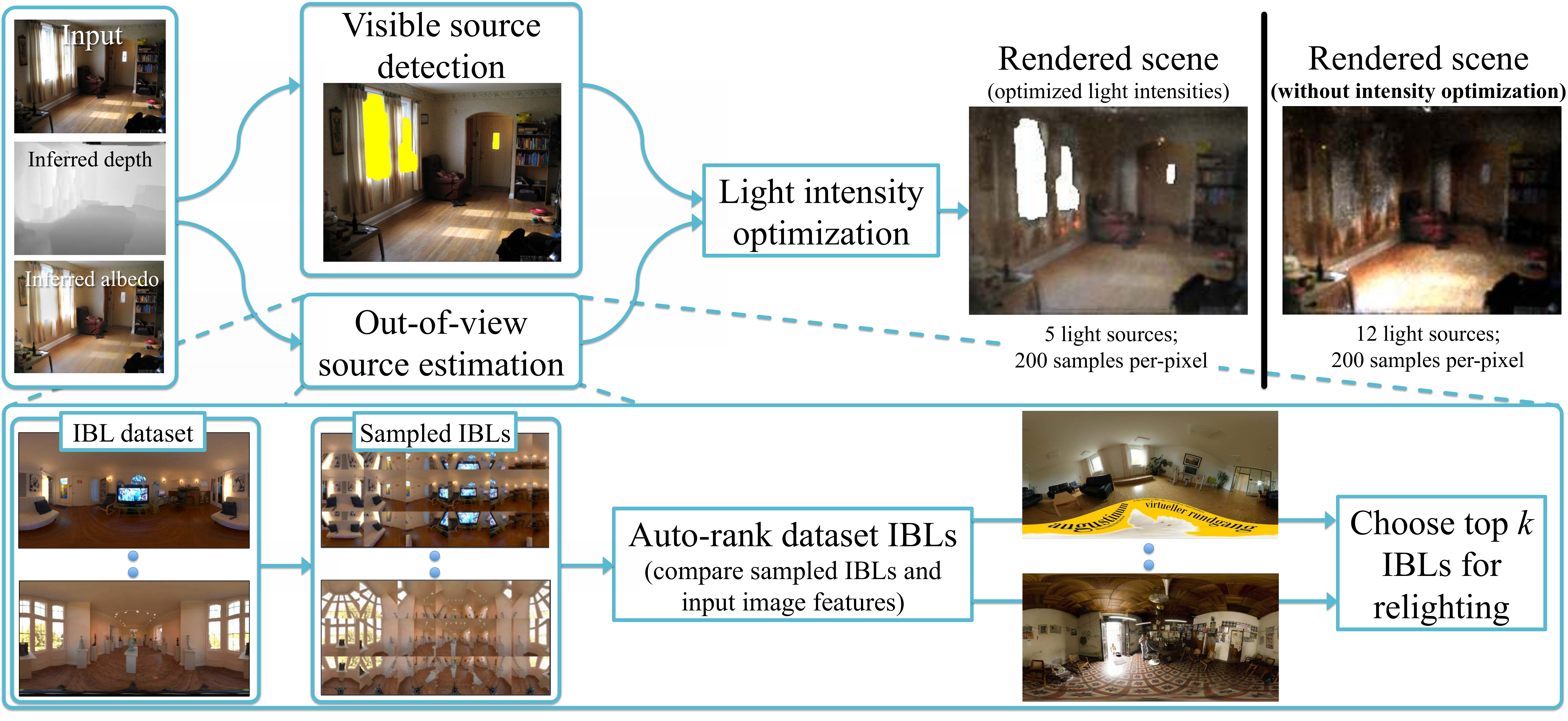}}
\vspace{-3mm}
\caption{Overview of our lighting estimation procedure. Light sources are first detected in the input image using our light classifier. To estimate light outside of the view frustum, we use a data-driven approach (utilizing the SUN360 panorama dataset). We train a ranking function to rank IBLs according to how well they ``match'' the input image's illumination (see text for details), and use the top $k$ IBLs for relighting. Finally, source intensities are optimized to produce a rendering of the scene that closely matches the input image. Our optimization not only encourages more plausible lighting conditions, but also improves rendering speed by pruning inefficient light sources.
}
\label{fig:light_overview}
\end{figure*}
%----------------------------------------------- Light overview

We categorize luminaires into {\it visible} sources (sources visible in the photograph), and {\it out-of-view} sources (all other luminaires). Visible sources are detected in the image using a trained ``light classifier'' (Sec~\ref{sec:illum:inside}). Out-of-view sources are estimated through a data-driven procedure (Sec~\ref{sec:illum:outside}) using a large dataset of annotated spherical panoramas (SUN360~\cite{Xiao:CVPR:12}).  The resulting lighting model is a hybrid of area sources and spherical emitting sources. Finally, light source intensities are estimated using an optimization procedure which adjusts light intensities so that the rendered scene appears similar to the input image (Sec~\ref{sec:illum:render}). Figure~\ref{fig:light_overview} illustrates this procedure.

\boldhead{Dataset} We have annotated light sources in 100 indoor and 100 outdoor scenes from the SUN360 dataset. The annotations also include a discrete estimate of distance from the camera (``close'': 1-5m, ``medium'': 5-50m, ``far'': $>$50m, ``infinite'': reserved for sun).  Since SUN360 images are LDR and tonemapped, we make no attempt to annotate absolute intensity, only the position/direction of sources. Furthermore, the goal of our classifier is only to predict location (not intensity). This data is used in both our in-view and out-of-view techniques below. 

%\boldhead{Parameterization} Our estimated illumination models are parameterized by emitting polygons (detected light sources in the photo, and the matched light sources from the out-of-view algorithm) and non-emitting regions (the parts of the photo not estimated to be sources, as well as the estimated LDR panorama for out-of-view lighting -- this mainly acts as a reflection map).

\subsection{Illumination visible in the view frustum}
\label{sec:illum:inside}
To detect sources of light in the image, we have developed a new light classifier. For a given image, we segment the image into superpixels using SLIC~\cite{Achanta:ijcv:12}, and compute features for each superpixel. We use the following features: the height of the superpixel in the image (obtained by averaging the 2D location of all pixels in the superpixel), as well as the features used by Make3D\footnote{17 edge/smoothing filter responses in YCbCr space are averaged over the superpixel. Both the energy and kurtosis of the filter responses are computed (second and fourth powers)  for a total of 34 features, and then concatenated with four neighboring (top,left,bottom,right) superpixel features. This is done at two scales (50\% and 100\%), resulting in $340 (=34\times 5 \times 2$) features per superpixel.} ~\cite{Saxena:pami09}. 

Using our annotated training data, we train a binary classifier to predict whether or not a superpixel is emitting/reflecting a significant amount of light using these features. For this task, we do not use the discrete distance annotations (these are however used in Sec~\ref{sec:illum:outside}). A classification result is shown in Figure~\ref{fig:light_overview}. In supplemental material, we show many more qualitative and quantitative results of our classifier, and demonstrate that our classifier significantly outperforms baseline detectors (namely thresholding).

For each detected source superpixel (and corresponding pixels), we find their 3D position using the pixel's estimated depth and the projection operator $K$. Writing $D$ as the estimated depth map and $(x,y)$ as a pixel's position in the image plane, the 3D position of the pixel is given by: 
\begin{equation}
\ve{X} = D(x,y) K^{-1} [x,y,1]^T,
\end{equation}
where $K$ is the intrinsic camera parameters (projection matrix) obtained in Section~\ref{sec:recon}. We obtain a polygonal representation of each light source by fitting a 3D quadrilateral to each cluster (oriented in the direction of least variance). Notice that this only provides the position of the visible light sources; we describe how we estimate intensity in Section~\ref{sec:illum:render}. Figure~\ref{fig:light_overview} (top) shows a result of our light detector.

One might wonder why we train using equirectangular images and test using rectilinear images. Dror et al.~\shortcite{dror2004sta} showed that many image statistics computed on equirectangular images follow the same distributions as those computed on rectilinear images; thus features computed in either domain should be roughly the same. We have also tested our method on both kinds of images (more results in the supplemental file), and see no noticeable differences.

\subsection{Illumination outside of the view frustum}
\label{sec:illum:outside}

Estimating lighting from behind the camera is arguably the most difficult task in single-image illumination estimation. We use a data-driven approach, utilizing the extensive SUN360 panorama dataset of Xiao et al.~\shortcite{Xiao:CVPR:12}. However, since this dataset is not available in HDR, we have annotated the dataset to include light source positions and distances.

Our primary assumption is that if two photographs have similar appearance, then the illumination environment beyond the photographed region will be similar as well. There is some empirical evidence that this can be true (e.g. recent image-extrapolation methods~\cite{Zhang:CVPR:13}), and studies suggest that people hallucinate out-of-frame image data by combining photographic evidence with recent memories~\cite{Intraub:JEP:1989}. 

Using this intuition, we develop a novel procedure for matching images to luminaire-annotated panoramas in order to predict out-of-view illumination. We sample each IBL into $N$ rectilinear projections (2D) at different points on the sphere and at varying fields-of-view, and match these projections to the input image using a variety of features described below (in our work, $N=10$ stratified random samples on the sphere with azimuth $\in [0, 2\pi)$, elevation $\in [-\frac{\pi}{6},\frac{\pi}{6}]$, FOV $\in [\frac{\pi}{3},\frac{\pi}{2}]$). See the bottom of Fig~\ref{fig:light_overview} for an illustration. These projections represent the images a camera with that certain orientation and field of view would have captured for that particular scene. By matching the input image to these projections, we can effectively "extrapolate" the scene outside the field of view and estimate the out-of-view illumination.

Given an input image, our goal is to find IBLs in our dataset that emulate the input's illumination. Our rectilinear sampled IBLs provide us with ground truth data for training: for each sampled image, we know the corresponding illumination. Based on the past success of rank prediction for data-driven geometry estimation~\cite{satkin_bmvc2012}, we use this data and train an IBL rank predictor (for an input image, rank the dataset IBLs from best to worst).

\boldhead{Features} After sampling the panoramas into rectilinear images, we compute seven features for each image: geometric context~\cite{Hoiem:iccv:2005}, orientation maps~\cite{Lee:CVPR09}, spatial pyramids~\cite{Lazebnik:cvpr:06}, HSV histograms (three features total), and the output of our light classifier (Sec~\ref{sec:illum:inside}). 

We are interested in ranking pairs of images, so our final feature describes how well two particular images match in feature space. The result is a 7-dimensional vector where each component describes the similarity of a particular feature (normalized to [0,1], where higher values indicate higher similarity). Similarity is measured using the histogram intersection score for histogram features (spatial pyramid -- sp -- and HSV -- h,s,v -- histograms), and following Satkin et al.~\shortcite{satkin_bmvc2012}, a normalized, per-pixel dot product (averaged over the image) is used for for other features (geometric contact -- gc, orientation maps -- om, light classifier -- lc).  

More formally, let $F_i, F_j$ be the features computed on images $i$ and $j$, and $x^i_j$ be the vector that measures similarity between the features/images:
\begin{align}
x^i_j = [&nd(F_i^{\text{gc}},F_j^{\text{gc}}), nd(F_i^{\text{om}}, F_j^{\text{om}}), nd(F_i^{\text{lc}}, F_j^{\text{lc}}), \nonumber \\ 
&hi(F_i^{\text{sp}}, F_j^{\text{sp}}), hi(F_i^{\text{h}}, F_j^{\text{h}}), hi(F_i^{\text{s}}, F_j^{\text{s}}), hi(F_i^{\text{v}}, F_j^{\text{v}})]^T,
\label{eq:features}
\end{align}
where $nd(\cdot), hi(\cdot)$ are normalized dot product and histogram intersection operators respectively. In order to compute per-pixel dot products, images must be the same size. To compute features on two images with unequal dimension, we downsample the larger image to have the same dimension as the smaller image.

\boldhead{Training loss metric} 
In order to discriminate between different IBLs, we need to define a distance metric that measure how similar one IBL is to another. We use this distance metric as the loss for our rank training optimization where it encodes the additional margin when learning the ranking function (Eq~\ref{eq:ranking}).

A na\"{\i}ve way to measure the similarity between two IBLs is to use pixel-wise or template matching~\cite{Xiao:CVPR:12}. However, this is not ideal for our purposes since it requires accurate correspondences between elements of the scene that may not be important from a relighting aspect. Since our primary goal is re-rendering, we define our metric on that basis of how different a set of canonical objects appear when they are illuminated by these IBLs. In particular, we render nine objects with varying materials (diffuse, glossy, metal, etc.) into each IBL environment (see supplemental material). Then, we define the distance as the mean L2 error between the renderings.

One caveat is that our IBL dataset isn't HDR, and we don't know the intensities of the annotated sources. So, we compute error as the minimum over all possible light intensities. Define $\mathbf{I}_i$ and $\mathbf{I}_j$ as two IBLs in our dataset, and $I_i=[I_i^{(1)},\ldots,I_i^{(n)}], I_j=[I_j^{(1)},\ldots,I_j^{(m)}]$ as column-vectorized images rendered by the IBLs for each of the IBLs' sources (here $\mathbf{I}_i$ has $n$ sources, and $\mathbf{I}_j$ has $m$). Since a change in the intensity of the IBL corresponds to a change of a scale factor for the rendered image, we define the distance as the minimum rendered error over all possible intensities ($y_i$ and $y_j$):
\begin{equation}
d(\mathbf{I}_i,\mathbf{I}_j) = \min_{y_i, y_j} || I_i y_i -I_j y_j ||, \text{ s.t. } ||[y_i^T,y_j^T] || = 1.
\label{eq:metric}
\end{equation}
The constraint is employed to avoid the trivial solution, and we solve this using SVD.

\boldhead{Training the ranking function}
Our goal is to train a ranking function that can properly rank images (and their corresponding IBLs) by how well their features match the input's. Let $w$ be a linear ranking function, and $x^i_j$ be features computed between images $i$ and $j$. Given an input image $i$ and any two images from our dataset (sampled from the panoramas) $j$ and $k$, we wish to find $w$ such that $w^T x^i_j > w^T x^i_k$ when the illumination of $j$ matches the illumination of $i$ better than the illumination of $k$.

To solve this problem, we perform a standard 1-slack, linear SVM-ranking optimization~\cite{Joachims:kdd:06}:
\begin{equation}
\argmin_{w,\xi} ||w||^2 + C \xi, \ \text{ s.t. } w^T x^i_j \geq w^T x^i_k + \delta^i_{j,k}  - \xi, \ \xi\geq 0,
\label{eq:ranking}
\end{equation}
where $x$ are pairwise image similarity features (Eq~\ref{eq:features}), and $\delta^i_{j,k} = \max(d(\mathbf{I}_i,\mathbf{I}_k)-d(\mathbf{I}_i,\mathbf{I}_j),0)$ is a hinge loss to encourage additional margin for examples with unequal distances (according to Eq~\ref{eq:metric}, where $\mathbf{I}_i$ is the IBL corresponding to image $i$).

\boldhead{Inference} To predict the illumination of a novel input image ($i$), we compute the similarity feature vector (Eq~\ref{eq:features}) for all input-training image pairs ($x^i_j, \ \forall j$), and sort the prediction function responses ($w^Tx^i_j$) in decreasing order. Then, we use choose the top $k$ IBLs (in our work, we use $k=1$ for indoor images, and $k=3$ outdoors to improve the odds of predicting the correct sun location). Figure~\ref{fig:light_overview} (bottom) shows one indoor result using our method (where $k=2$ for demonstration).

\subsection{Intensity estimation through rendering}
\label{sec:illum:render}
Having estimated the location of light sources within and outside the image, we must now recover the relative intensities of the sources. Given the exact geometry and material of the scene (including light source positions), we can estimate the intensities of the sources by adjusting them them until a rendered version of the scene matches the original image. While we do not know exact geometry/materials, we assume that our automatic estimates are good enough, and could apply the above rendering-based optimization to recover the light source intensities; Fig~\ref{fig:light_overview} shows this process. Our optimization has two goals: match the rendered image to the input, and differing from past techniques~\cite{Boivin:2001,Karsch:SA2011}, ensure the scene renders efficiently.

Define $L_i$ as the intensity of the $i^{th}$ light source, $I$ as the input image, and $R(\cdot)$ as a scene ``rendering'' function that takes a set of light intensities and produces an image of the scene illuminated by those lights (i.e., $R(L)$). We want to find the intensity of each light source by matching the input and rendered images, so we could solve $\argmin_{L} \sum_{i \in \text{pixels}} ||I_i - R_i(L)||$. However, this optimization can be grossly inefficient and unstable, as it requires a new image to be rendered for each function evaluation, and rendering in general is non-differentiable. However, we can use the fact that light is additive, and write $R(\cdot)$ as a linear combination of ``basis'' renders~\cite{Schoeneman:1993,Nimeroff:94}. We render the scene (using our estimated geometry and diffuse materials) using only one light source at a time (i.e., $L_k = 1, L_j = 0 \ \forall j \neq k$, implying $L = e_k$). This results in one rendered image per light source, and we can write a new render function as $R'(w) = C \left(\sum_k w_k R(e_k)\right)$, where $C$ is the camera response function, and $R(e_i)$ is the scene rendered with only the $i^{th}$ source. In our work, we assume the camera response can be modeled as an exponent, i.e., $C(x) = x^\gamma$.  This allows us to rewrite the matching term above as
\begin{equation}
Q(w,\gamma) = \sum_{i \in \text{pixels}} \left| \left| I_i - \left[\sum_{k \in \text{sources}} w_k R_i(e_k)\right]^\gamma \right| \right|.
\label{eq:light_opt2}
\end{equation}
Since each basis render, $R(e_k)$, can be precomputed prior to the optimization, $Q$ can be minimized more efficiently than the originally described optimization.

We have hypothesized a number of light source locations in Secs~\ref{sec:illum:inside} and~\ref{sec:illum:outside}, and because our scene materials are purely diffuse, and our geometry consists of a small set of surface normals, there may exist an infinite number of lighting configurations that produce the same rendered image. Interestingly, user studies have shown that humans cannot distinguish between a range of illumination configurations~\cite{Ramanarayanan07:VisualEq}, suggesting that there is a family of lighting conditions that produce the same perceptual response. This is actually advantageous, because it allows our optimization to choose only a small number of ``good'' light sources and prune the rest. In particular, since our final goal is to relight the scene with the estimated illumination, we are interested in lighting configurations that can be rendered faster. We can easily detect if a particular light source will cause long render times by rendering the image with the given source for a fixed amount of time, and checking the variance in the rendered image (i.e., noise); we incorporate this into our optimization. By rendering with fewer sources and sources that contribute less variance, the scenes produced by our method render significantly faster than without this optimization (see Fig~\ref{fig:light_overview}, right).

Specifically, we ensure that only a small number of sources are used in the final lighting solution, and also prune problematic light sources that may cause inefficient rendering. We encode this with a sparsity prior on the source intensities and a smoothness prior on the rendered images: 
\begin{equation}
P(w) = \sum_{k \in \text{sources}} \left[ ||w_k||_1 + w_k \sum_{i \in \text{pixels}} || \nabla R_i(e_k) || \right].
\label{eq:light_priors}
\end{equation}
Intuitively, the first term coerces only a small number of nonzero elements in $w$, and the second term discourages noisy basis renders from having high weights (noise in a basis render typically indicates an implausible lighting configuration, making the given image render much more slowly).

%----------------------------------------------- User interface
\begin{figure*}
\setlength\fboxrule{1pt}
\setlength\fboxsep{0pt}
\begin{overpic}[width=.24\linewidth,clip,trim=0 0 114 0]{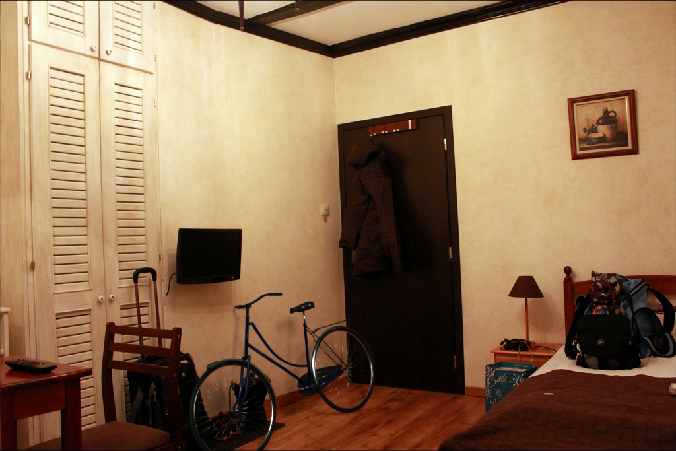}
\put(0,55){\fcolorbox{white}{white}{\includegraphics[width=.10\linewidth,clip,trim=0 0 114 0]{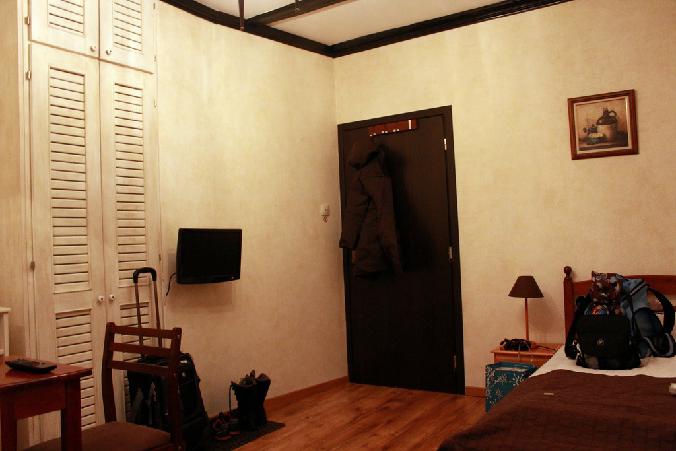}}} 
\setlength\fboxsep{2pt}
\put(0,2){\colorbox{white}{\Large (a)}}
\end{overpic}
\setlength\fboxsep{0pt}
\begin{overpic}[width=.24\linewidth,clip,trim=0 0 114 0]{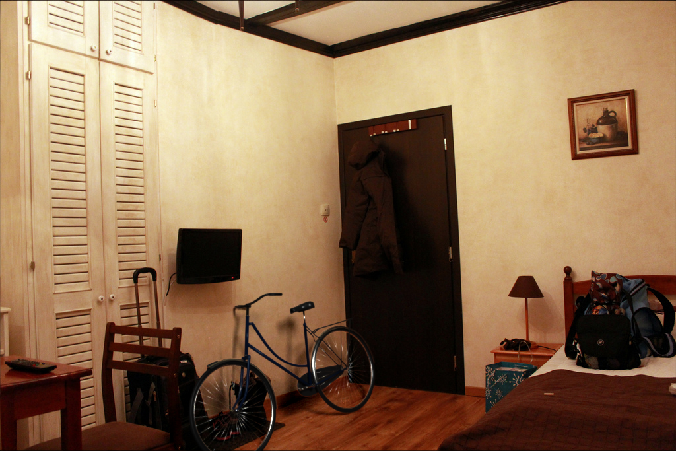} 
\setlength\fboxsep{2pt}
\put(0,2){\colorbox{white}{\Large (b)}}
\end{overpic}
\hfill %
\setlength\fboxsep{0pt}
\begin{overpic}[width=.24\linewidth]{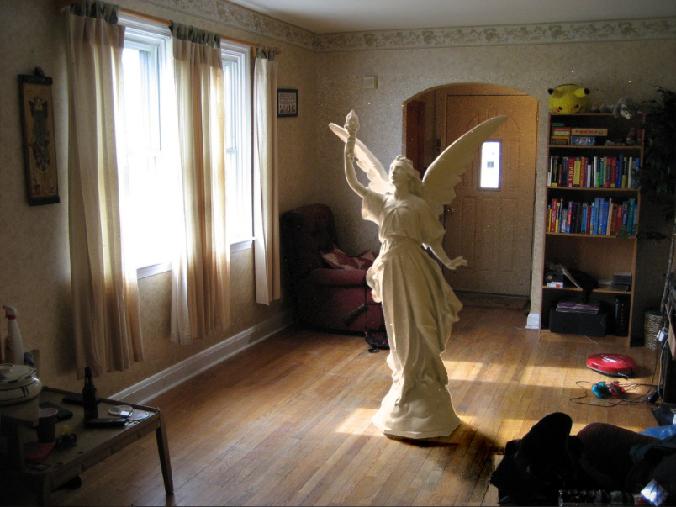}
\put(0,55){\fcolorbox{white}{white}{\includegraphics[width=.10\linewidth]{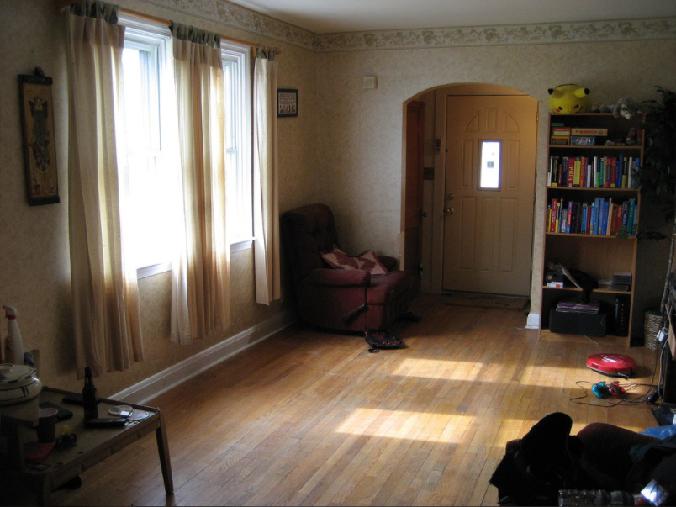}}} 
\setlength\fboxsep{2pt}
\put(0,2){\colorbox{white}{\Large (c)}}
\end{overpic}
\setlength\fboxsep{0pt}
\begin{overpic}[width=.24\linewidth]{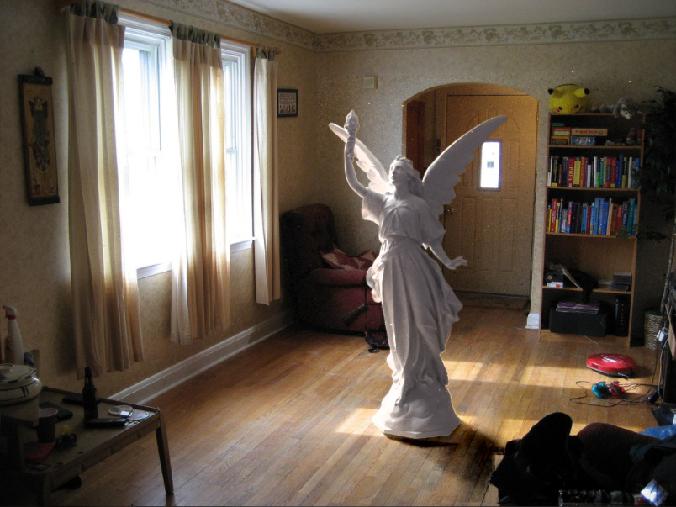} 
\setlength\fboxsep{2pt}
\put(0,2){\colorbox{white}{\Large (d)}}
\end{overpic}
\vspace{1mm} \\
\setlength\fboxsep{0pt}
\begin{overpic}[width=.33\linewidth]{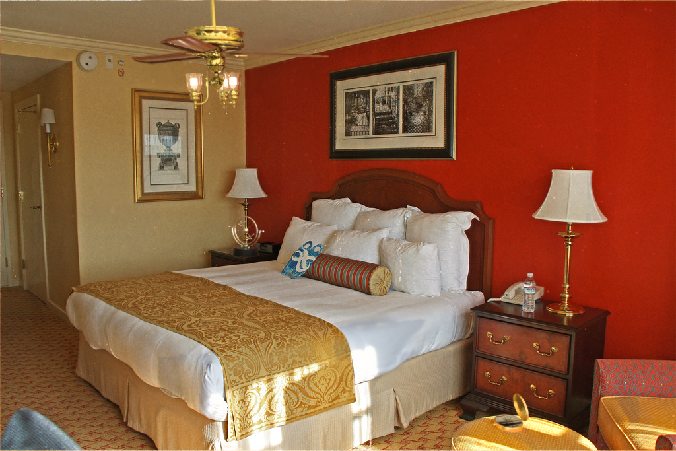}
\put(100,67){\fcolorbox{white}{white}{\includegraphics[width=.13\linewidth]{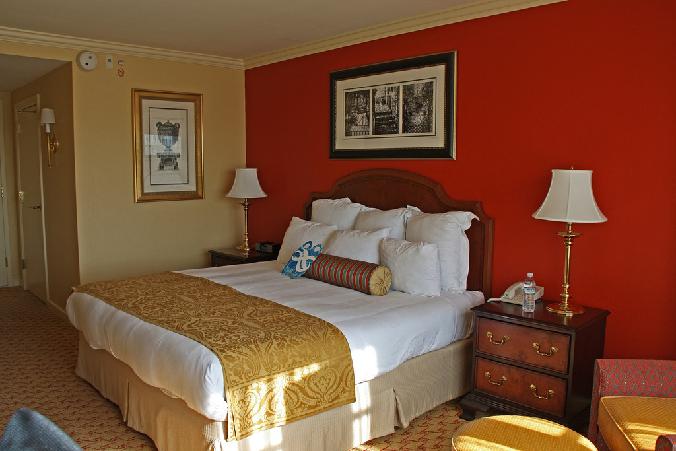}}}
\end{overpic}
\setlength\fboxsep{2pt}
\begin{overpic}[width=.33\linewidth]{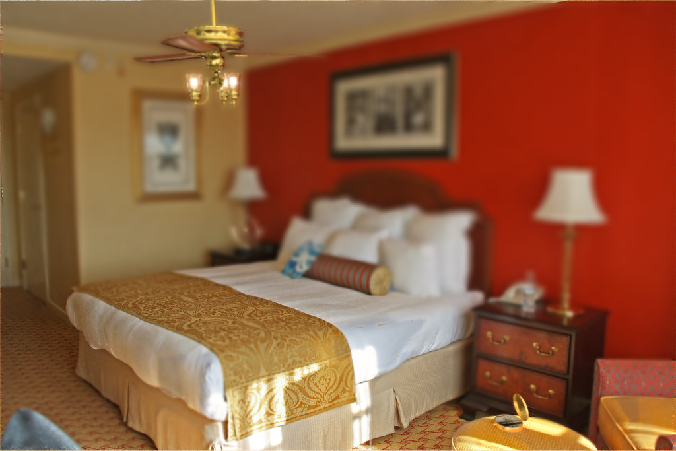} \put(0,2){\colorbox{white}{\Large (e)}}\end{overpic}
\setlength\fboxsep{2pt}
\begin{overpic}[width=.33\linewidth]{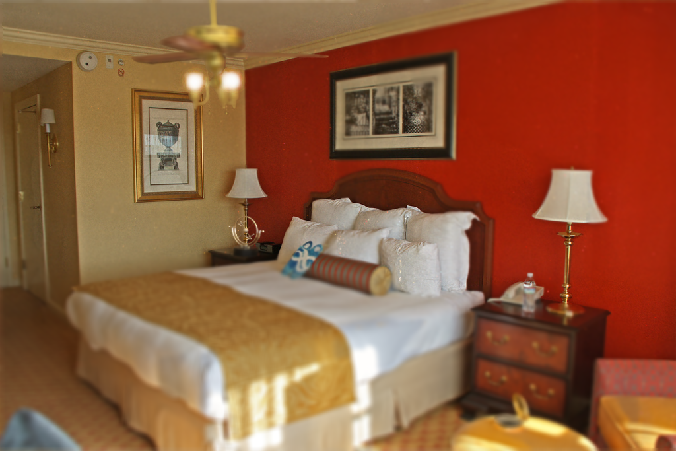} \put(0,2){\colorbox{white}{\Large (f)}}\end{overpic}
\caption{Our user interface provides a drag-and-drop mechanism for inserting 3D models into an image (input image overlaid on our initial result). Our system also allows for real-time illumination changes without re-rendering. Using simple slider controls, shadows and interreflections can be softened or hardened (a,b), light intensity can be adjusted to achieve a different mood (c,d), and synthetic depth-of-field can be applied to re-focus the image (e,f). See the accompanying video for a full demonstration. Best viewed in color at high-resolution. {\it Photo credits: {\footnotesize \textcopyright{}}Rachel Titiriga (top) and Flickr user {\footnotesize \textcopyright{}}``Mr.TinDC'' (bottom).}
}
\label{fig:interface}
\end{figure*}
%----------------------------------------------- User interface 

Combining the previous equations, we develop the following optimization problem:
\begin{align}
\argmin_{w,\gamma} \ \ Q(w,\gamma) + \lambda_{P} P(w) + \lambda_{\gamma} ||\gamma-\gamma_0||, \nonumber \\
\text{s.t. } w_k \geq 0 \ \forall k,\ \gamma>0,
\label{eq:light_opt3}
\end{align}
where $\lambda_{P} = \lambda_{\gamma} = 0.1$ are weights, and $\gamma_0=\frac{1}{2.2}$. We use a continuous approximation to the absolute value ($|x| \approx \sqrt{x^2+\epsilon}$), and solve using the active set algorithm~\cite{Nocedal2006NO}. The computed weights ($w$) can be directly translated into light intensities ($L$), and we now have an entire model of the scene (geometry, camera, and materials from Sec~\ref{sec:recon}, and light source positions/intensities as described above). 

Our method has several advantages to past ``optimization-through-rendering'' techniques~\cite{Karsch:SA2011}. First, our technique has the ability to discard unnecessary and inefficient light sources by adding illumination priors (Eq~\ref{eq:light_priors}). Second, we estimate the camera response function jointly during our optimization, which we do not believe has been done previously. 
%, and could be used as a separate application (e.g., camera response from a single image). 
Finally, by solving for a simple linear combination of pre-rendered images, our optimization procedure is much faster than previous methods that render the image for each function evaluation (e.g., as in~\cite{Karsch:SA2011}). Furthermore, our basis lights could be refined or optimized with user-driven techniques guided by aesthetic principles (e.g.~\cite{Boyadzhiev:SG:2013}).

%%%%%%%%%%%%%%%%%%%%%%%%%%%%%%%%%%%%%%%%%%%%%%%%
\section{Physically grounded image editing}
\label{sec:interface}

One of the potential applications of our automatically generated 3D scene models is physically grounded photo editing. In other words, we can use the approximate scene models to facilitate physically-based image edits. For example, one can use the 3D scene to insert and composite objects with realistic lighting into a photograph, or even adjust the depth-of-field and aperture as a post-process.  

There are many possible interactions that become available with our scene model, and we have developed a prototype interface to facilitate a few of these. Realistically inserting synthetic objects into legacy photographs is our primary focus, but our application also allows for post-process lighting and depth-of-field changes. To fully leverage our scene models, we require physically based rendering software. We use LuxRender\footnote{\url{http://www.luxrender.net}}, and have built our application on top of LuxRender's existing interface. Figure~\ref{fig:interface} illustrates the possible uses of the interface, and we refer the reader to the accompanying video for a full demonstration. 

\boldhead{Drag-and-drop insertion} Once a user specifies an input image for editing, our application automatically computes the 3D scene as described in sections~\ref{sec:recon} and~\ref{sec:illum}. Based on the inserted location, we also add additional geometric constraints so that the depth is flat in a small region region around the base of the inserted object. For a given image of size $1024\times768$, our method takes approximately five minutes on a 2.8Ghz dual core laptop to estimate the 3D scene model, including depth and illumination (this can also be precomputed or computed remotely for efficiency, and only occurs once per image). Next, a user specifies a 3D model and places it in the scene by simply clicking and dragging in the picture (as in Fig~\ref{fig:teaser}). Our scene reconstruction facilitates this insertion: our estimated perspective camera ensures that the object is scaled properly as it moves closer/farther from the camera, and based on the surface orientation of the clicked location, the application automatically re-orients the inserted model so that its up vector is aligned with the surface normal. Rigid transformations are also possible through mouse and keyboard input (scroll to scale, right-click to rotate, etc).

Once the user is satisfied with the object placement, the object is rendered into the image\footnote{Rendering time is clearly dependent on a number of factors (image size, spatial hierarchy, inserted materials, etc), and is slowed by the fact that we use an unbiased ray-tracer. The results in this paper took between 5 minutes and several hours to render, but this could be sped up depending on the application and resources available.}. We use the additive differential rendering method~\cite{Debevec:98} to composite the rendered object into the photograph, but other methods for one-shot rendering could be used (e.g. Zang et al.~\shortcite{Zang:2012}). This method renders two images: one containing synthetic objects $I_{obj}$, and one without synthetic objects $I_{noobj}$, as well as an object mask M (scalar image that is 0 everywhere where no object is present, and (0, 1] otherwise). The final composite image $I_{final}$ is obtained by
\begin{equation}
I_{final} = M \odot I_{obj} + (1-M) \odot (I_{b}+I_{obj}-I_{noobj}),
\label{eq:composite}
\end{equation}
where $I_{b}$ is the input image, and $\odot$ is the entry-wise product. For efficiency (less variance and overhead), we have implemented this equation as a surface integrator plugin for LuxRender. Specifically, we modify LuxRender's bidirectional path tracing implementation~\cite{PBRTv2} so that pixels in $I_{final}$ are intelligently sampled from either the input image or the rendered scene such that inserted objects and their lighting contributions are rendered seamlessly into the photograph. We set the maximum number of eye and light bounces to 16, and use LuxRender's default Russian Roulette strategy (dynamic thresholds based on past samples).

The user is completely abstracted from the compositing process, and only sees $I_{final}$ as the object is being rendered. The above insertion method also works for adding light sources (for example, inserting and emitting object). Figure~\ref{fig:teaser} demonstrates a drag-and-drop result.

\boldhead{Lighting adjustments} Because we have estimated sources for the scene, we can modify the intensity of these sources to either add or subtract light from the image. Consider the compositing process (Eq~\ref{eq:composite} with no inserted objects; that is, $I_{obj} = I_{noobj}$, implying $I_{final} = I_{b}$. Now, imagine that the intensity of a light source in $I_{obj}$ is increased (but the corresponding light source in $I_{noobj}$ remains the same).  $I_{obj}$ will then be brighter than $I_{noobj}$, and the compositing equation will reflect a physical change in brightness (effectively rendering a more intense light into the scene).  Similarly, decreasing the intensity of source(s) in $I_{obj}$ will remove light from the image. Notice that this is a physical change to the lighting in the picture rather than tonal adjustments (e.g., applying a function to an entire image). 

This technique works similarly when there are inserted objects present in the scene, and a user can also manually adjust the light intensities in a scene to achieve a desired effect (Fig~\ref{fig:interface}, top). Rather than just adjusting source intensities in $I_{obj}$, if sources in both $I_{obj}$ and $I_{noobj}$ are modified equally, then only the intensity of the inserted object and its interreflections (shadows, caustics, etc) will be changed (i.e. without adding or removing light from the rest of the scene).

By keeping track of the contribution of each light source to the scene, we only need to render the scene once, and lighting adjustments can be made in real time without re-rendering (similar to~\cite{Loscos:1999,Gibson:2000}). Figure~\ref{fig:interface} shows post-process lighting changes; more can be found in supplemental material.

\boldhead{Synthetic depth-of-field} Our prototype also supports post-focusing the input image: the user specifies a depth-of-field and an aperture size, and the image is adaptively blurred using our predicted depth map ($\ve{D}$). Write $I_{final}$ as the composite image with inserted objects (or the input image if nothing is inserted), and $G(\sigma)$ as a Gaussian kernel with standard deviation $\sigma$. We compute the blur at the $i^{th}$ pixel as
\begin{equation} 
I_{dof,i} = I_{final,i} \star G( a|\ve{D}_i -  d| ),  
\end{equation}
where $d$ is the depth of field, and $a$ corresponds to the aperture size. Figures~\ref{fig:teaser} and~\ref{fig:interface} (bottom) shows a post-focus result. 

%%%%%%%%%%%%%%%%%%%%%%%%%%%%%%%%%%%%%%%%%%%%%%%%
\section{Evaluation}
\label{sec:study}

We conducted two user studies to evaluate the ``realism'' achieved by our insertion technique. Each user study is a series of two-alternative forced choice tests, where the subject chooses between a pair of images which he/she feels looks the most realistic.

In the first study (Sec~\ref{sec:study:real}), we use real pictures. Each pair of images contains one actual photograph with a small object placed in the scene, and the other is photo showing a similar object inserted synthetically into the scene (without the actual object present).

The second study (Sec~\ref{sec:study:synthetic}) is very similar, but we use highly realistic, synthetic images rather than real pictures. For one image, a synthetic object is placed into the full 3D environment and rendered (using unbiased, physically-based ray tracing); in the other, the same 3D scene is rendered without the synthetic object (using the same rendering method), and then synthetically inserted {\it into the photograph} (rather than the 3D scene) using our method.

Finally, we visualize the quantitative accuracy achieved by our estimates in Section~\ref{sec:study:quant}.

\subsection{Real image user study}
\label{sec:study:real}
\boldhead{Experimental setup}  We recruited 30 subjects for this study, and we used a total of 10 synthetic objects across three unique, real scenes. Each subject saw each inserted object only once, and viewed seven to nine side-by-side trial images; one was a real photograph containing a real object, and the other is a synthetic image produced by our method (containing a synthetic version of the real object in the real photo). Subjects were asked to choose the image they felt looked most realistic. Trials and conditions were permuted to ensure even coverage. An example trial pair is shown in Fig~\ref{fig:userstudy}, and more can be found in supplemental material.

\boldhead{Conditions} We tested six binary conditions that we hypothesized may contribute to a subject's performance: 
{\bf expert} subject; {\bf realistic shape}; {\bf complex material}; {\bf automatic} result; {\bf multiple objects} inserted; and whether the trial occured in the {\bf first half} of a subject's study (to test whether there was any learning effect present). Subjects were deemed experts if they had graphics/vision experience (in the study, this meant graduate students and researchers at a company lab), and non-experts otherwise. The authors classified objects into realistic/synthetic shape and complex/simple material beforehand. We created results with our automatic drag-and-drop procedure as one condition (``automatic''), and using our interface, created an improved (as judged by the authors) version of each result by manually adjusting the light intensities (``refined''). 

\boldhead{Results} On average, the result generated by our {\bf automatic}  method was selected as the real image in 34.1\% of the 232 pairs that subjects viewed. The refined condition achieved 35.8\% confusion, however these distributions do not significantly differ using a two-tailed t-test. An optimal result would be 50\%. For comparison, Karsch et al.~\shortcite{Karsch:SA2011} evaluated their semiautomatic method with a similar user study, and achieved 34\% confusion, although their study (scenes, subjects, etc) were not identical to ours.

We also tested to see if any of the other conditions (or combination of conditions) were good predictors of a person's ability to perform this task. We first used logistic regression to predict which judgements would be correct using the features, followed by a Lasso ($L_1$ regularized logistic regression~\cite{Tibshirani}, in the Matlab R2012b implementation) to identify features that predicted whether the subject judged a particular pair correctly.   We evaluated deviance (mean negative log-likelihood on held out data) with 10-fold cross validation.  The lowest  deviance regression ignores all features; incorporating the ``expert'' and ``realistic shape'' features causes about  a 1\% increase in the deviance; incorporating other features causes the deviance to increase somewhat.  This strongly suggests that none of the features actually affect performance.   

\begin{figure}[t]
\begin{overpic}[width=.49\linewidth,clip,trim=120 60 60 60]{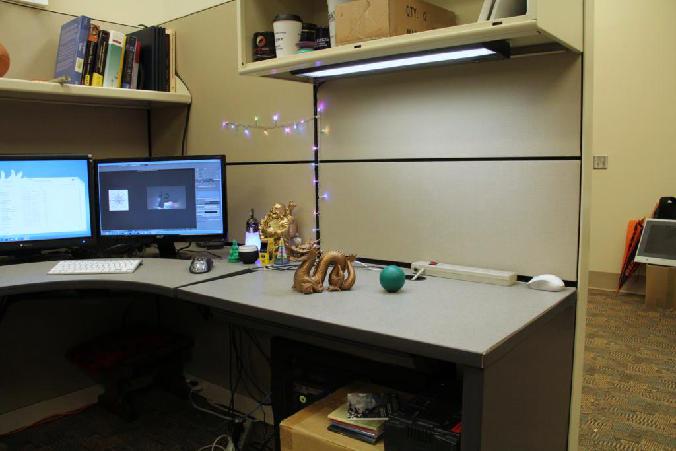}\put(0,0){\colorbox{white}{\large A1}}
\end{overpic}
\begin{overpic}[width=.49\linewidth,clip,trim=120 60 60 60]{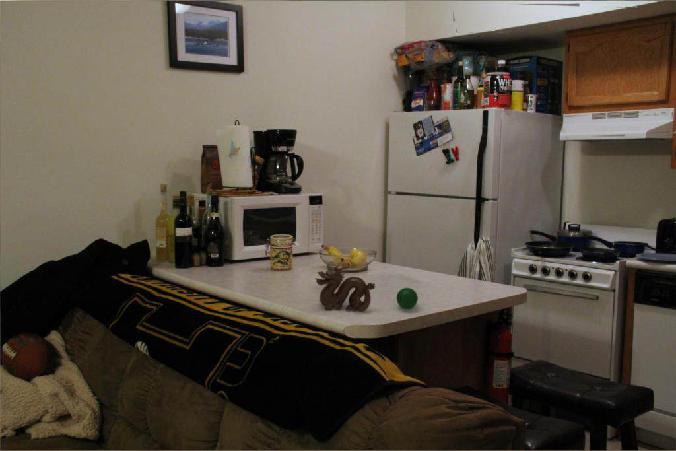}\put(0,0){\colorbox{white}{\large A2}}
\end{overpic}
\vspace{1mm} \\
\begin{overpic}[width=.49\linewidth,clip,trim=0 0 0 0]{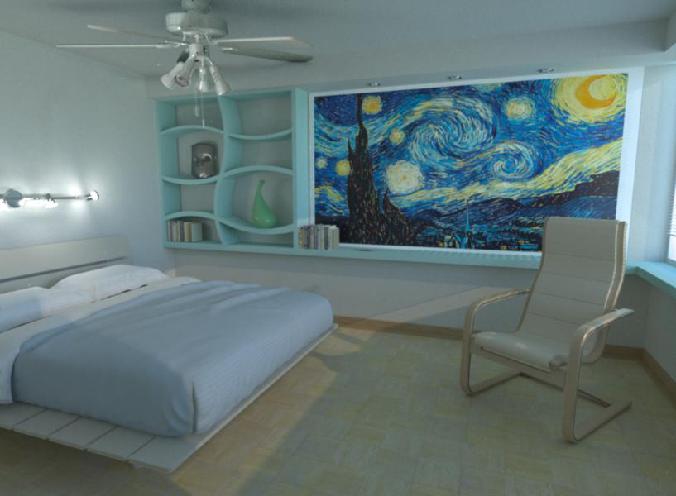}\put(0,2){\colorbox{white}{\large B1}}
\end{overpic}
\begin{overpic}[width=.49\linewidth,clip,trim=0 0 0 0]{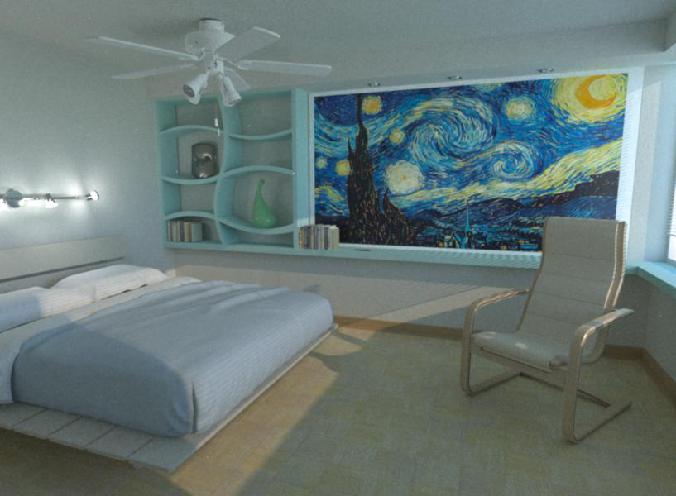}\put(0,2){\colorbox{white}{\large B2}}
\end{overpic}
\caption{Example trials from our ``real image'' user study (top) and our ``synthetic image'' user study (bottom). In the studies, users were shown two side-by-side pictures; one photograph is real (or rendered photorealistically with an exact 3D scene), and the other has synthetic objects inserted into it with our method. Users were instructed to choose the picture from the pair that looked the most realistic. For each row, which of the pair would you choose? Best viewed in color at high-resolution.  {\it Scene modeling credits (bottom): Matthew Harwood.}
\protect{\begin{turn}{180} \centerline{\small \it A1,B2: real/fully rendered; A2, B1: created with our method}\end{turn}}
}
\label{fig:userstudy}
\end{figure}
%----------------------------------------------- User study 

\subsection{Synthetic image user study}
\label{sec:study:synthetic}
Our ``real image'' study provides encouraging initial results, but it is limited due to the nature of data collection. 
Collecting corresponding real and synthetic objects with high-quality geometry and reflectance properties can be extremely difficult, so we confined our real image study to small, tabletop accessories (household items and 3D-printable items).

In this follow-up study, we utilize highly realistic, synthetic 3D scenes in order to more extensively evaluate our method (objects with larger-pixel coverage and varying materials/geometry, diverse lighting conditions, indoors/outdoors).

\boldhead{Experimental setup} We collected four synthetic (yet highly realistic\footnote{We pre qualified these images as highly-confusable with real pictures in a preliminary study; see supplemental material.}) scenes -- three indoor, one outdoor. For each scene, we inserted three realistic 3D models using conventional modeling software (Blender, \url{http://www.blender.org/}), and rendered each scene using LuxRender under three lighting conditions (varying from strongly directed to diffuse light), for a total of 36 (= 4 scenes $\times$ 3 objects $\times$ 3 lighting conditions) unique and varied scenes (viewable in supplemental material). Next, we used our method to insert the same 3D models in roughly the same location into the empty, rendered images, resulting in 36 ``synthetic'' insertion results corresponding to the 36 ground truth rendered images. 

For the study, each participant viewed 12 pairs\footnote{Image pairs are randomly permuted/selected so that each of the 12 objects appears exactly once to each subject.} of corresponding images, and was asked to select which image he/she felt looked the most realistic (two-alternative forced choice). For example, a subject might see two identical bedroom scenes with a ceiling fan, except in one picture, the fan had actually been inserted using our method (see Fig~\ref{fig:userstudy}).

We polled 450 subjects using Mechanical Turk. In an attempt to avoid inattentive subjects, each study also included four ``qualification'' image pairs (a cartoon picture next to a real image) placed throughout the study in a stratified fashion. Subjects who incorrectly chose any of the four cartoon picture as realistic were removed from our findings (16 in total, leaving 434 studies with usable data).

At the end of the study, we showed subjects two additional image pairs: a pair containing rendered spheres (one a physically plausible, the other not), and a pair containing line drawings of a scene (one with proper vanishing point perspective, the other not). For each pair, subjects chose the image they felt looked most realistic. Then, each subject completed a brief questionnaire, listing demographics, expertise, and voluntary comments.

%----------------------------------------------- User study results (conditions)
\begin{table}
\tbl{Fraction of times subjects chose a synthetic insertion result over the ground truth insertion in our user study}{
\begin{tabular}{| l c || c | c | c |} \hline
Condition & N & ours & Khan & Lalonde/matching \\ \hline

indoor & 1332 & \cellcolor{darkyellow} {\bf .377} & .311 & .279 \\
outdoor & 444 &  \cellcolor{darkyellow} .288 & {\bf .297} & .240 \\  \hline

diffuse light & 1184 &  \cellcolor{darkyellow} {\bf .377} & \cellcolor{darkyellow}  .332 &  \cellcolor{darkyellow} .238 \\ 
directional light & 592 &  \cellcolor{darkyellow} {\bf .311} &  \cellcolor{darkyellow} .257 & \cellcolor{darkyellow}  .330 \\ \hline

simple material & 1036 & {\bf .353} & .301 & \cellcolor{darkyellow}.248 \\
complex material & 740 & {\bf .357  }  &    .315     &  \cellcolor{darkyellow} .299\\  \hline

large coverage  & 1332 & {\bf .355} & \cellcolor{darkyellow} .324 & .277 \\
small coverage  & 444 & {\bf .356} & \cellcolor{darkyellow}  .253 & .245 \\  \hline

%near geometry?  & 888 & ..360 & .324 & .279  \\
%far from geom? & 888 & .349 & .290 & .260\\  \hline

good composition  & 1036 & {\bf .356} & .312 & .265 \\
poor composition  & 740 & {\bf .353} & .300 & .274 \\  \hline

good perspective  & 1036 & \cellcolor{darkyellow}  {\bf .378} & .316 & .287 \\
poor perspective  & 740 & \cellcolor{darkyellow}  {\bf .322} & .295 & .244 \\  \hline

male  & 1032 & {\bf .346} & .295 & .267 \\
female  & 744 & {\bf .367} & .319 & .272 \\  \hline

age ($\leq$25)  & 468 & {\bf .331} &  .294  & \cellcolor{darkyellow}   .234 \\
age ($>$25) & 1164 &  {\bf .363} & .312 & \cellcolor{darkyellow} .284 \\  \hline

color normal  & 1764 & {\bf .353} & .308 & .267 \\
not color normal & 12 & {\bf .583} & .167 & .361 \\  \hline

passed p-s tests & 1608 & \cellcolor{darkyellow}  {\bf .341} & \cellcolor{darkyellow}  .297 & \cellcolor{darkyellow}  .260 \\
failed p-s tests & 168 &  \cellcolor{darkyellow}  .482 & \cellcolor{darkyellow}   {\bf .500} & \cellcolor{darkyellow}  .351 \\  \hline 

non-expert  & 1392 & {\bf .361} & .312 & .275 \\
expert  & 384 & {\bf .333} & .288 & .252 \\  \hline 
\hline
overall & 1776 &  {\bf .355} & .307 & .269  \\ \hline

\end{tabular}
}
\vspace{2mm}
\label{tab:study_results_conditions}
\begin{tabnote}
Highlighted blocks indicate that there are significant differences in confusion when a particular condition is on/off ($p$ value $< 0.05$ using a 2-tailed t-test). For example, in the top left cell, the ``ours indoor'' distribution was found to be significantly different from the ``ours outdoor'' distribution. For the Lalonde/matching column, the method Lalonde et al. is used for outdoor images, and the template matching technique is used indoors (see text for details). The best method for each condition is shown in bold. N is the total number of samples. Overall, our confusion rate of 35.5\% better than the baselines (30.7\% and 26.9\%) by statistically significant margins.
\end{tabnote}
\end{table}
%----------------------------------------------- User study results (conditions)

\boldhead{Methods tested} We are primarily interested in determining how well our method compares to the ground truth images, but also test other illumination estimation methods as baselines.  We generate results using the method of Khan et al.~\shortcite{Khan:2006:IME:1179352.1141937} (projecting the input image onto a hemisphere, duplicating it, and using this as the illumination environment), the method of Lalonde et al.~\shortcite{Lalonde:2009vp} (for outdoor images only), and a simplified IBL-matching technique that finds a similar IBL to the input image by template matching (similar to~\cite{Xiao:CVPR:12}; we use this method indoors only).

For each method, only the illumination estimation stage of our pipeline changes (depth, reflectance, and camera estimation remain the same), and all methods utilize our lighting optimization technique to estimate source intensity (Sec~\ref{sec:illum:render}).

\boldhead{Conditions} Because we found no condition in our initial study to be a useful predictor of people's ability to choose the real image, we introduce many new (and perhaps more telling) conditions in this study: {\bf indoor/outdoor} scene, {\bf diffuse/direct lighting} in the scene, {\bf simple/complex material} of inserted object, {\bf good/poor composition} of inserted object, and if the inserted object has {\bf good/poor perspective}. We also assigned each subject a set of binary conditions: {\bf male/female}, {\bf age $<25$ / $\geq 25$}, {\bf color-normal / not color-normal}, whether or not the subject correctly identified both the physically accurate sphere {\it and} the proper-perspective line drawing at the end of the study ({\bf passed/failed perspective-shading (p-s) tests}), and also {\bf expert/non-expert} (subjects were classified as experts only if they passed the perspective-shading tests {\it and} indicated that they had expertise in art/graphics).

\boldhead{Results and discussion}
Overall, our synthetic image study showed that people confused our insertion result with the true rendered image in over 35\% of 1776 viewed image pairs (an optimal result would be 50\%). We also achieve better confusion rates than the methods we compared to, and Table~\ref{tab:study_results_conditions} shows these results. 

Although the differences between our method and the other methods might appear smaller ($\scriptstyle\mathtt{\sim}$5-10 percentage points), these differences are statistically significant using a two-tailed test. Furthermore, we are only assessing differences in light estimation among these method (since it is our primary technical contribution); every other part of our pipeline remains constant (e.g. camera and depth estimation, as well as the light intensity optimization). In other words, we do not compare directly to other techniques, rather these other lighting techniques are aided by bootstrapping them with the rest of our pipeline.

We also analyze which conditions lead to significant differences in confusion for each method, indicated by highlighted cells in the table. As expected, our method works best indoors and when the lighting is not strongly directed, but still performs reasonably well otherwise (over 31\% confusion). Also, our method looks much more realistic when inserted objects have reasonable perspective.

Perhaps most interestingly, our method is almost confusable at chance with ground truth for people who have a hard time detecting shading/perspective problems in photographs (failed p-s tests), and the population that passed is significantly better at this task than those that failed. This could mean that such a test is actually a very good predictor for whether a person will do well at this task, or it could be viewed as a secondary ``qualification'' test (i.e. similar yet possibly more difficult than the original cartoon-real image qualification tests). Either way, for the population that passed these tests, the confusion rates are very similar to the overall mean (i.e. results are very similar even if this data is disregarded).

All other conditions did not have a significant impact on the confusion rate for our method. Again, we observe that the ``expert'' subjects did not perform significantly better ($p$ value = 0.323) than ``non-expert'' subjects (by our classification), although they did choose the ground truth image more times on average.

\subsection{Ground truth comparison}
\label{sec:study:quant}

\begin{figure}
\begin{center}
\begin{minipage}{1.0\linewidth}
\begin{center}
\begin{minipage}{0.325\linewidth} 
\centerline{Ground truth} \vspace{0.5mm}
\includegraphics[width=\linewidth,clip=true,trim=0 40 0 130]{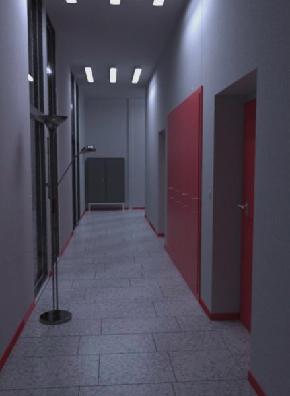}\\
\includegraphics[width=\linewidth,clip=true,trim=0 200 250 0]{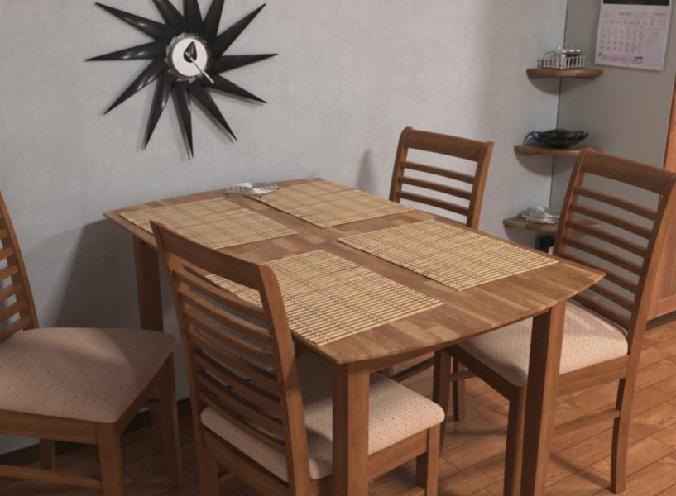}\\
\includegraphics[width=\linewidth]{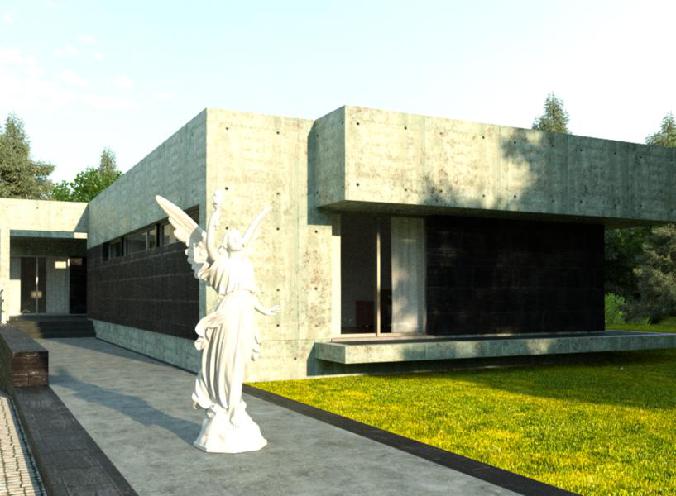}
\end{minipage}
\begin{minipage}{0.325\linewidth} 
\centerline{Our result} \vspace{0.5mm}
\includegraphics[width=\linewidth,clip=true,trim=0 40 0 130]{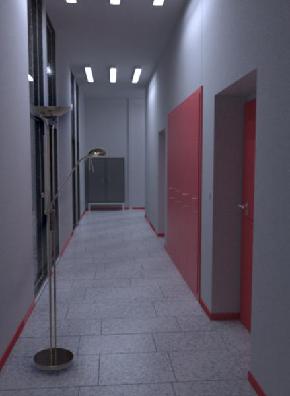}\\
\includegraphics[width=\linewidth,clip=true,trim=0 200 250 0]{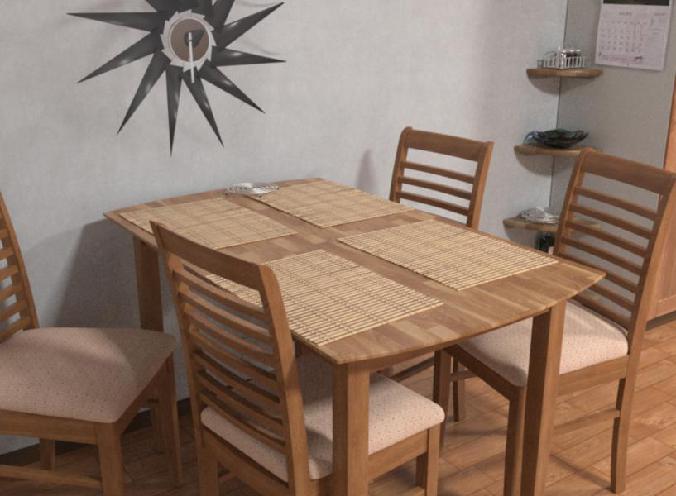}\\
\includegraphics[width=\linewidth]{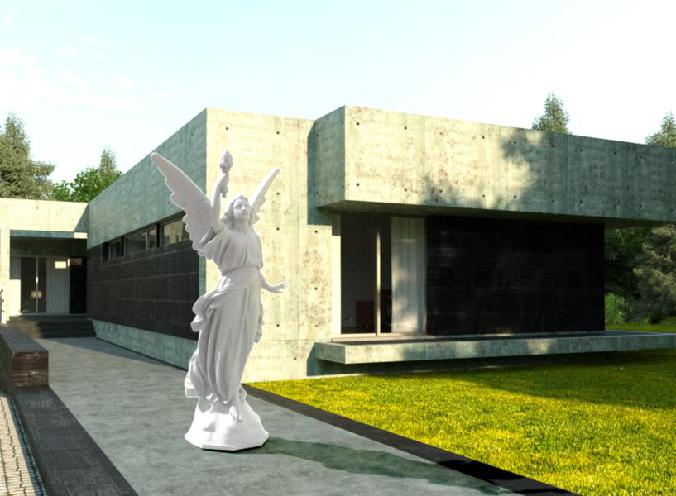}
\end{minipage}
\begin{minipage}{0.325\linewidth} 
\centerline{\cite{Khan:2006:IME:1179352.1141937} lighting}
\includegraphics[width=\linewidth,clip=true,trim=0 40 0 130]{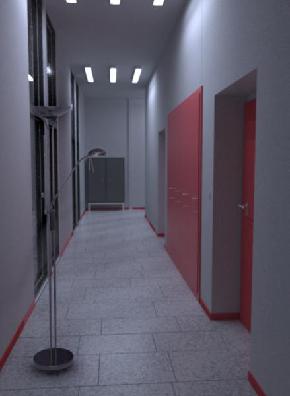}\\
\includegraphics[width=\linewidth,clip=true,trim=0 200 250 0]{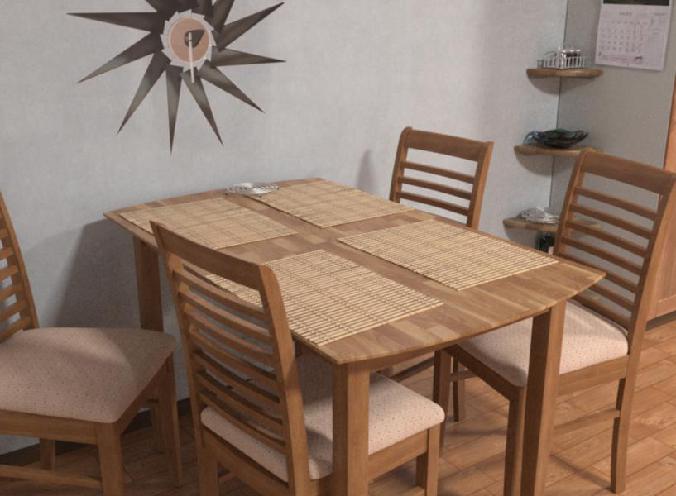}\\
\includegraphics[width=\linewidth]{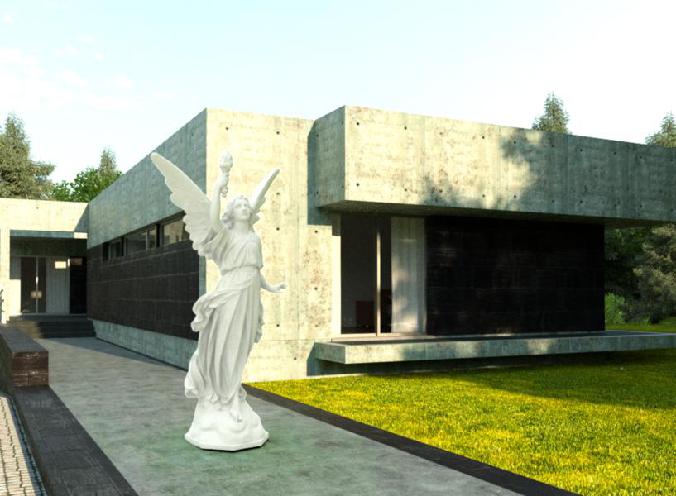}
\end{minipage}
\end{center}
\end{minipage}
\end{center}
\caption{Example results used in our synthetic user study (corresponding to the scenes shown in Figure~\protect\ref{fig:groundtruth}). We compared the ground truth renderings (left) to our insertion method (middle) and others (e.g. using the method of Khan et al. for relighting, right). {\it Scene modeling credits (top to bottom): {\footnotesize \textcopyright{}}Simon Wendsche, {\footnotesize \textcopyright{}}Peter Sandbacka, {\footnotesize \textcopyright{}}Eduardo Camara.}
}
\label{fig:studysynth_results}
\end{figure}

We also measure the accuracy of our scene estimates by comparing to ground truth. In Figure~\ref{fig:studysynth_results}, we show three images from our ``synthetic'' user study: rendered images (ground truth) are compared with insertion results using our technique as well as results relit using the method of Khan et al.~\shortcite{Khan:2006:IME:1179352.1141937} (but our method is used for all other components, e.g. depth, light optimization, and reflectance). 

Figure~\ref{fig:groundtruth} shows our inverse rendering estimates (depth, reflectance, and illumination) for the same three scenes as in Figure~\ref{fig:studysynth_results}. These images illustrate typical errors produced by our algorithm: depth maps can be confused by textured surfaces and non-planar regions, reflectance estimates may fail in the presence of hard shadows and non-Lambertian materials, and illumination maps can contain inaccurate lighting directions or may not appear similar to the actual light reflected onto the scene. These downfalls suggest future research directions for single image inference.

Notice that while in some cases our estimates differ greatly from ground truth, our insertion results can be quite convincing, as demonstrated by our user studies. This indicates that the absolute physical accuracy of inverse rendering is not a strong indicator of how realistic an inserted object will look; rather, we believe that relative cues are more important. For example, inserted objects typically cast better shadows in regions of planar geometry and where reflectance is constant and relatively correct (i.e. erroneous up to a scale factor); strong directional illuminants need not be perfectly located, but should be consistent with ambient light. Nonetheless, 3D object insertion will likely benefit from improved estimates.

For quantitative errors and additional results, we refer the reader to supplemental material.

%----------------------------------------------- Synthetic user study images
%----------------------------------------------- Synthetic user study images

%----------------------------------------------- Quantitative results
\newcommand{\includestudydecomposition}[8]{
\begin{center}
\begin{minipage}{#6\linewidth}
\begin{center}
\begin{minipage}{0.2\linewidth} 
\vspace{13mm}\centerline{\Large #1}\vspace{13mm} 
\centerline{Rendered image} 
\includegraphics[width=\linewidth,clip=true,trim=#2 #3 #4 #5]{fig/userstudy-synthetic/img/#1-#7.jpg}
\end{minipage}
\begin{minipage}{0.2\linewidth} 
\centerline{True depth}
\includegraphics[width=\linewidth,clip=true,trim=#2 #3 #4 #5]{fig/userstudy-synthetic/decompositions/depth/#1-true.jpg}\\
\centerline{Estimated depth} 
\includegraphics[width=\linewidth,clip=true,trim=#2 #3 #4 #5]{fig/userstudy-synthetic/decompositions/depth/#1-#7-ours.jpg}
\end{minipage}
\begin{minipage}{0.2\linewidth} 
\centerline{True reflectance} \vspace{0.5mm}
\includegraphics[width=\linewidth,clip=true,trim=#2 #3 #4 #5]{fig/userstudy-synthetic/decompositions/refl/#1-true.jpg}\\
\centerline{Estimated reflectance}\vspace{0.5mm}
\includegraphics[width=\linewidth,clip=true,trim=#2 #3 #4 #5]{fig/userstudy-synthetic/decompositions/refl/#1-#7-ours.jpg}
\end{minipage}
\begin{minipage}{#8\linewidth} 
\centerline{True illumination} \vspace{0.5mm}
\includegraphics[width=\linewidth]{fig/userstudy-synthetic/decompositions/illum/#1-#7-true.jpg}\\
\centerline{Estimated illumination}\vspace{0.5mm}
\includegraphics[width=\linewidth]{fig/userstudy-synthetic/decompositions/illum/#1-#7-ours.jpg}
\end{minipage}
\end{center}
\end{minipage}
\end{center}
}

\begin{figure*}
\includestudydecomposition{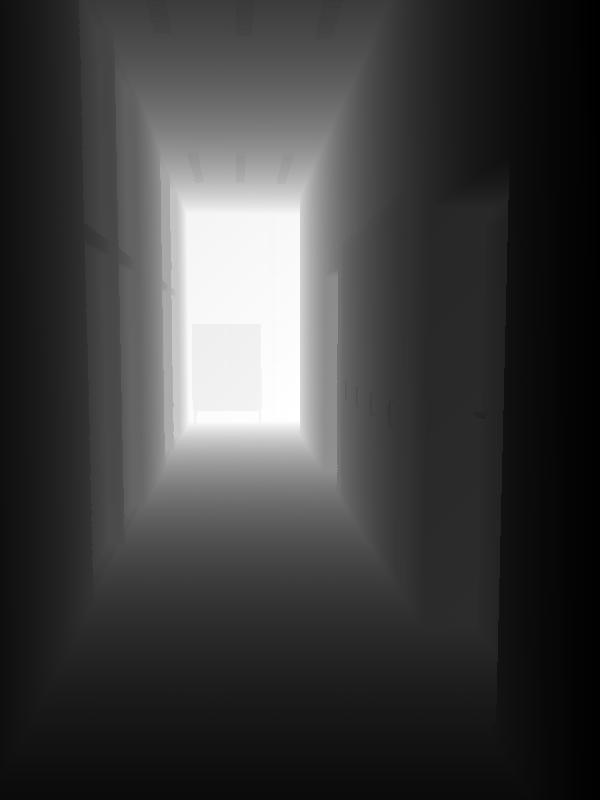}{0}{80}{0}{200}{0.85}{C}{0.345}
\vspace{3mm}
\centerline{\rule{0.95\linewidth}{0.1mm}}
\vspace{3mm}
\includestudydecomposition{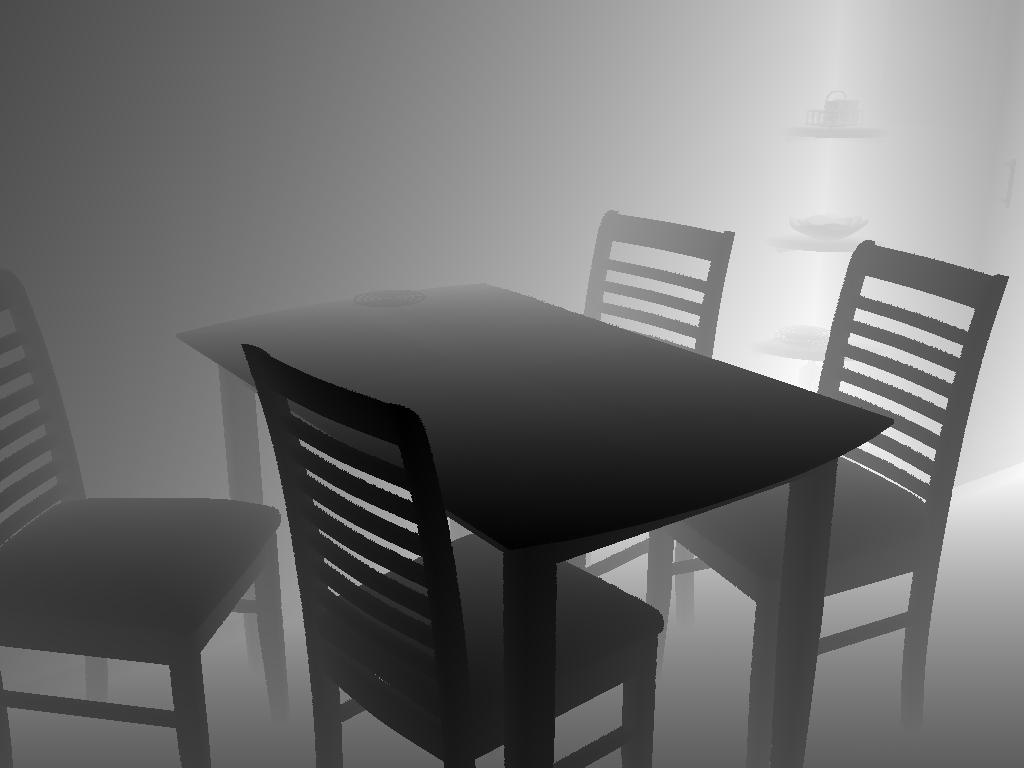}{0}{0}{0}{0}{1}{B}{0.295}
\vspace{3mm}
\centerline{\rule{0.95\linewidth}{0.1mm}}
\vspace{3mm}
\includestudydecomposition{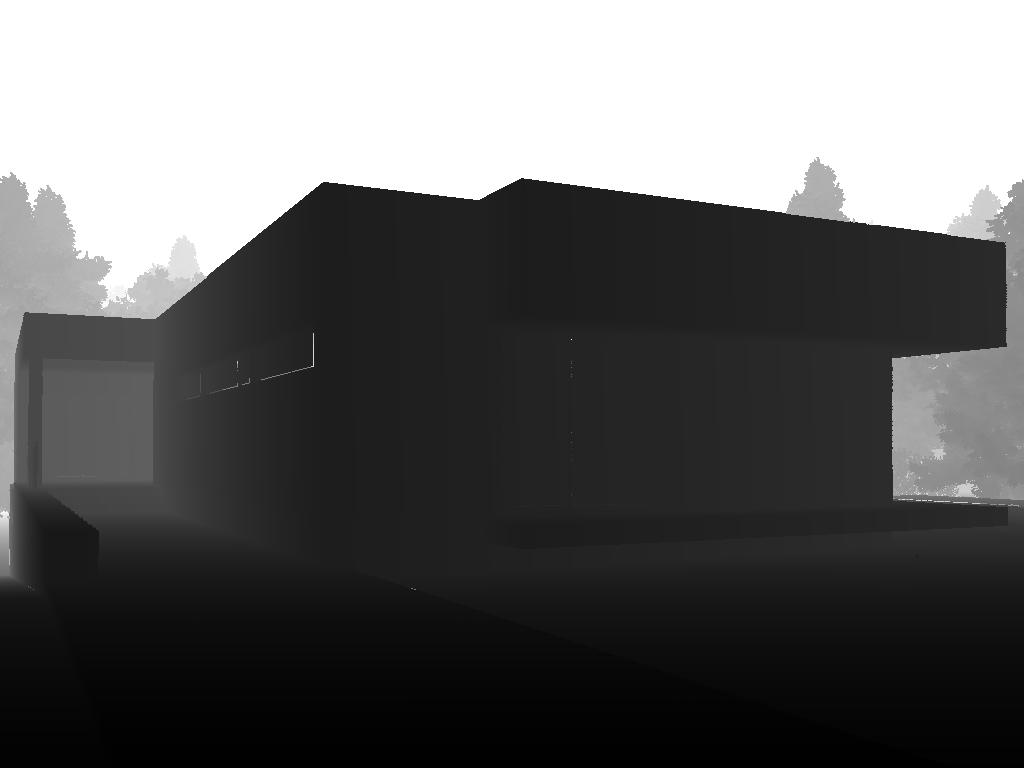}{0}{0}{0}{0}{1}{A}{0.295}
\caption{Comparison of ground truth scene components (depth, diffuse reflectance, and illumination) with our estimates of these components. Illumination maps are tonemapped for display. See Figure~\protect\ref{fig:studysynth_results} and the supplemental document for insertion results corresponding to these scenes. {\it Scene modeling credits (top to bottom): {\footnotesize \textcopyright{}}Simon Wendsche, {\footnotesize \textcopyright{}}Peter Sandbacka, {\footnotesize \textcopyright{}}Eduardo Camara.}}
\label{fig:groundtruth}
\end{figure*}
%----------------------------------------------- Quantitative results

%%%%%%%%%%%%%%%%%%%%%%%%%%%%%%%%%%%%%%%%%%%%%%%%
\section{Results and conclusion}
\label{sec:results}

%----------------------------------------------- Additional results
\begin{figure*}
\setlength\fboxsep{0pt}
\setlength\fboxrule{1pt}
\begin{centering}
\begin{minipage}{1\linewidth}
\begin{overpic}[width=.32\linewidth,clip=true,trim=0 50 0 0]{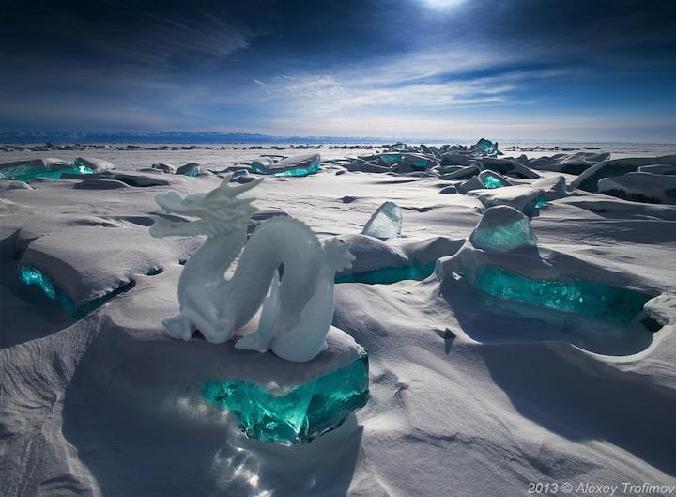}
\put(95,62){\fcolorbox{white}{white}{\includegraphics[width=0.13\linewidth,clip,trim=0 10 0 0]{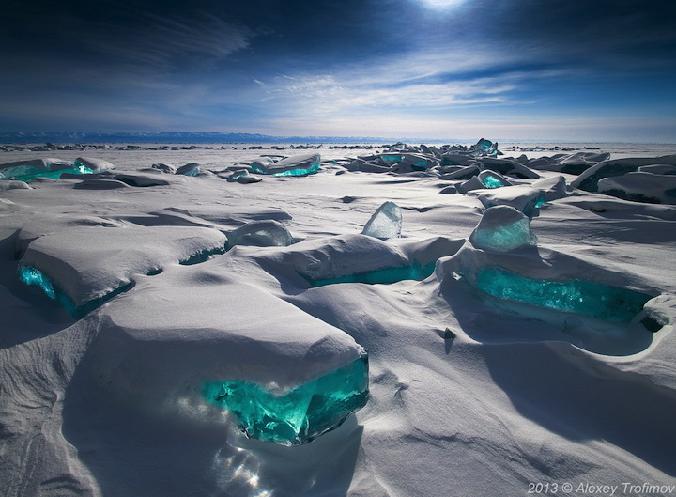}}}
\end{overpic}
\hfill
\begin{overpic}[width=.32\linewidth,clip=true,trim=0 20 0 72]{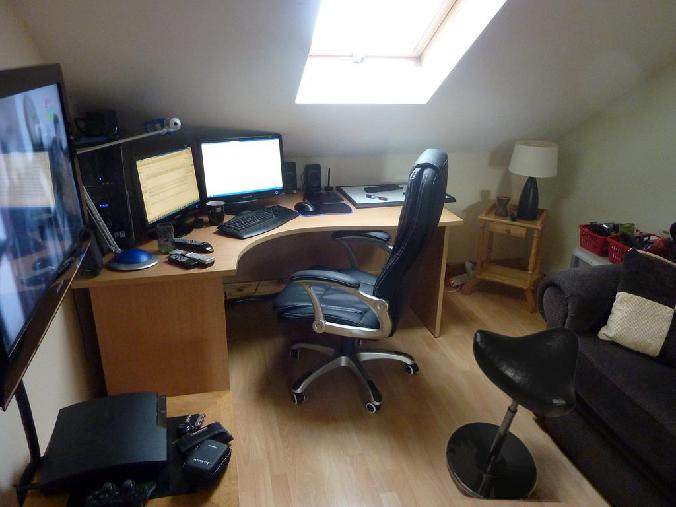}
\put(0,1){\fcolorbox{white}{white}{\includegraphics[width=0.13\linewidth,clip,trim=0 20 0 72]{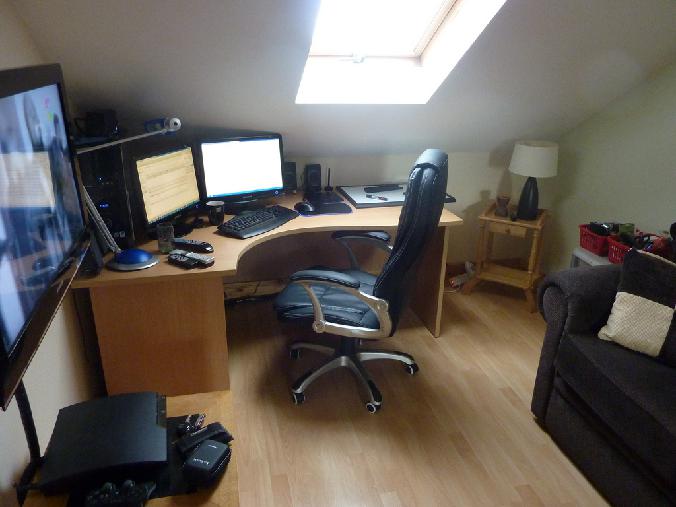}}}
\end{overpic}
\hfill
\begin{overpic}[width=.32\linewidth,clip=true,trim=0 0 0 110]{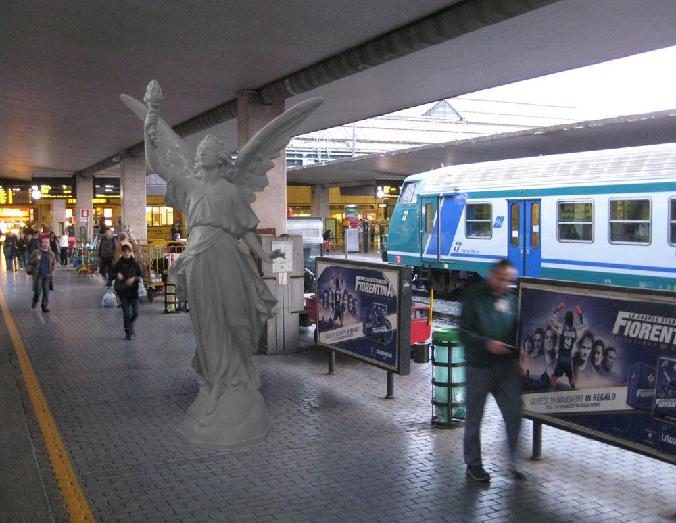}
\put(95,56){\fcolorbox{white}{white}{\includegraphics[width=0.13\linewidth,clip,trim=0 0 0 0]{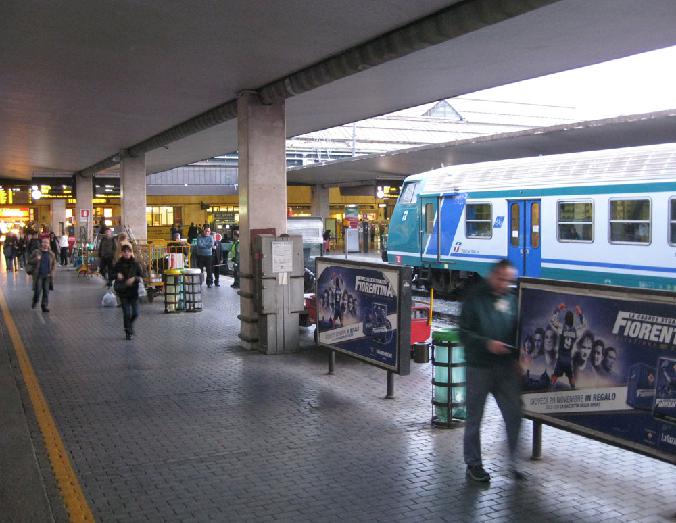}}}
\end{overpic}
\\
\begin{overpic}[width=.32\linewidth,clip=true,trim=0 0 0 25]{fig/results_supp/martini_lux.jpg}
\put(0,1){\fcolorbox{white}{white}{\includegraphics[width=0.13\linewidth,clip,trim=0 0 0 25]{fig/results_supp/martini.jpg}}}
\end{overpic}
\hfill
\begin{overpic}[width=.32\linewidth,clip,trim=18 18 18 188]{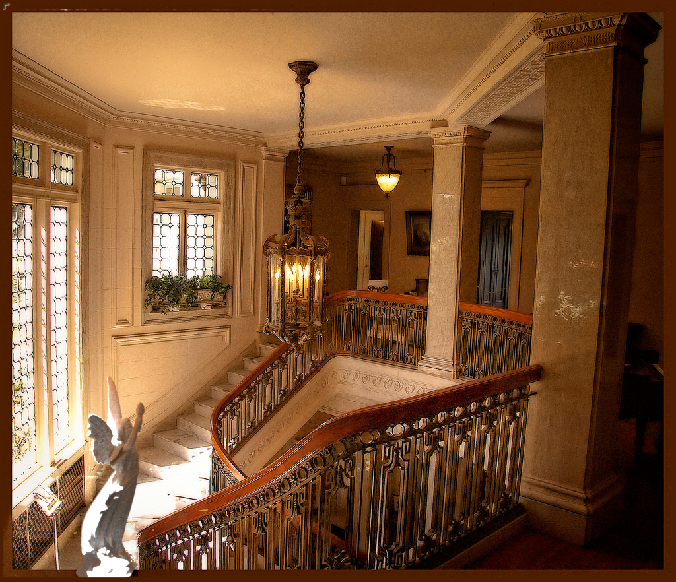}
\put(95,65){\fcolorbox{white}{white}{\includegraphics[width=.13\linewidth,clip,trim=18 18 18 188]{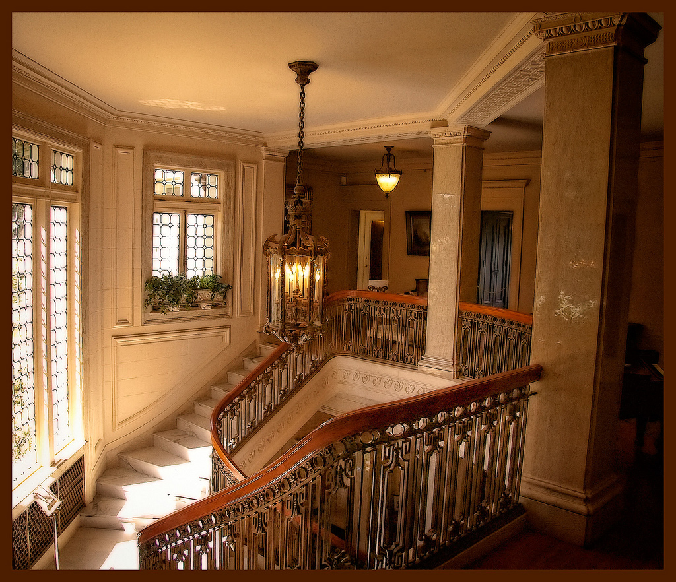}}}
\end{overpic}
\hfill
\begin{overpic}[width=.32\linewidth,clip,trim=67 18 0 18]{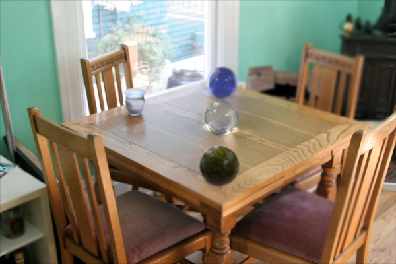}
\put(0,1){\fcolorbox{white}{white}{\includegraphics[width=.13\linewidth,clip,trim=67 18 0 18]{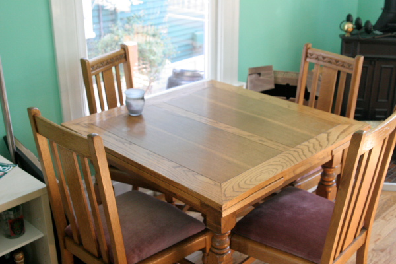}}}
\end{overpic}
\end{minipage}
\end{centering}
\vspace{-3mm}
\caption{Additional results produced by our method (original picture is overlaid on top of the result). Our method is capable of producing convincing results both indoors and outdoors, and can be robust to non-manhattan scenes. Our method can be applied to arbitrary images, and makes no explicit assumptions about the scene geometry; virtual staging is one natural application of our system. Best viewed in color at high-resolution. {\it Photo credits (left to right, top to bottom): {\footnotesize \textcopyright{}}Alexey Trofimov, {\footnotesize \textcopyright{}}Sean MacEntee, Flickr users {\footnotesize \textcopyright{}}``AroundTuscany,'' {\footnotesize \textcopyright{}}``Wonderlane,'' {\footnotesize \textcopyright{}}``PhotoAtelier,'' and {\footnotesize \textcopyright{}}Brian Teutsch}.
}
\label{fig:results}
\end{figure*}
%----------------------------------------------- Additional results 

We show typical results produced by our system in Fig~\ref{fig:results}. Our method is applicable both indoors and outdoors and for a variety of scenes. Although a Manhattan World is assumed at various stages of our algorithm, we show several reasonable results even when this assumption does not hold. Failure cases are demonstrated in Fig~\ref{fig:failure_examples}. Many additional results (varying in quality and scene type) can be found in the supplemental document. The reader is also referred to the accompanying video for a demonstration of our system.

As we found in our user study, our method is better suited for indoor scenes or when light is not strongly directional. In many cases, people confused our insertion results as real pictures over one third of the time. For outdoor scenes, we found that simpler illumination methods might suffice (e.g.~\cite{Khan:2006:IME:1179352.1141937}), although our geometry estimates are still useful for these scenes (to estimate light intensity and to act as shadow catchers).

We would like to explore additional uses for our scene models, such as for computer gaming and videos. It would be interesting to try other physically grounded editing operations as well; for example, deleting or moving objects from a scene, or adding physically-based animations when inserting objects (e.g., dragging a table cloth over a table). Extending our method to jointly infer a scene all at once (rather than serially) could lead to better estimates. Our depth estimates might also be useful for inferring depth order and occlusion boundaries.

Judging by our synthetic image user study, our method might also be useful as a fast, incremental renderer. For example, if a 3D modeler has created and rendered a scene, and wants to insert a new object quickly, our method could be used rather than re-rendering the full scene (which, for most scenes, should incur less render time since our estimated scene may contain significantly fewer light sources and/or polygons).

We have presented a novel image editor that, unlike most image editors that operate in 2D, allows users to make physically meaningful edits to an image with ease. Our software supports realistic object insertion,  on-the-fly lighting changes, post-process depth of field modifications, and is applicable to legacy, LDR images. These interactions are facilitated by our automatic scene inference algorithm, which encompasses our primary technical contributions: single image depth estimation, and data-driven illumination inference. Results produced by our system appear realistic in many cases, and a user study provides good evidence that objects inserted with our fully automatic technique look more realistic than corresponding real photographs over one third of the time.

%----------------------------------------------- Failure cases
\begin{figure*}
\setlength\fboxrule{1pt}
\setlength\fboxsep{0pt}
\begin{minipage}{.32\linewidth}
\begin{overpic}[width=\linewidth,clip=true,trim=0 0 0 0]{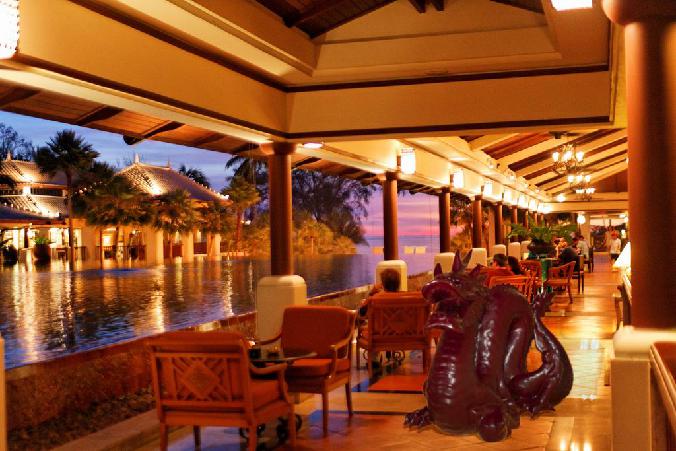}
\put(0,1){\fcolorbox{white}{white}{\includegraphics[width=0.33\linewidth,clip,trim=0 0 0 0]{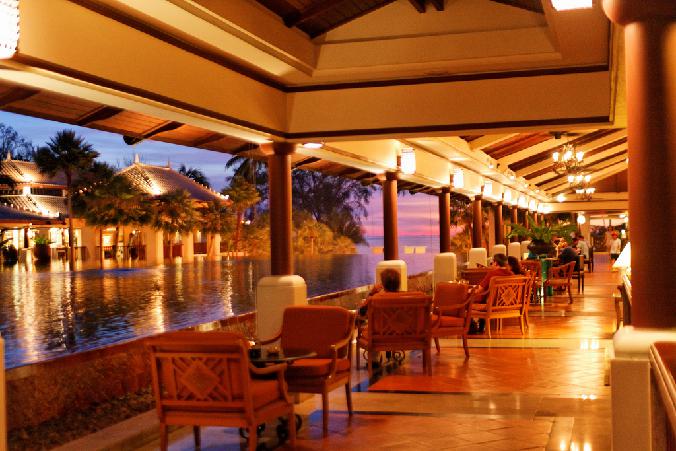}}}
\end{overpic}\\
\begin{overpic}[width=\linewidth,clip=true,trim=0 100 0 0]{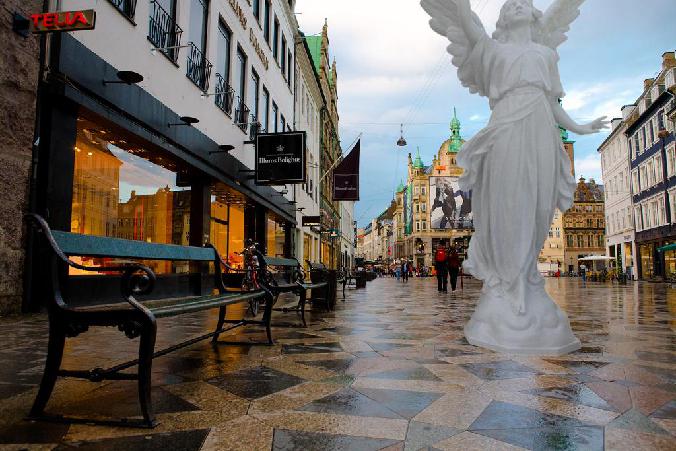}
\put(0,1){\fcolorbox{white}{white}{\includegraphics[width=0.33\linewidth,clip,trim=0 0 0 0]{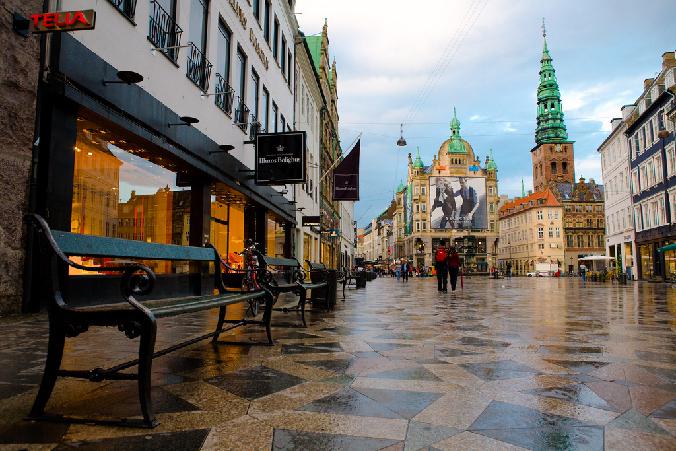}}}
\end{overpic}
\end{minipage}
\hfill
\begin{minipage}{.32\linewidth}
\begin{overpic}[width=\linewidth,clip=true,trim=0 65 0 0]{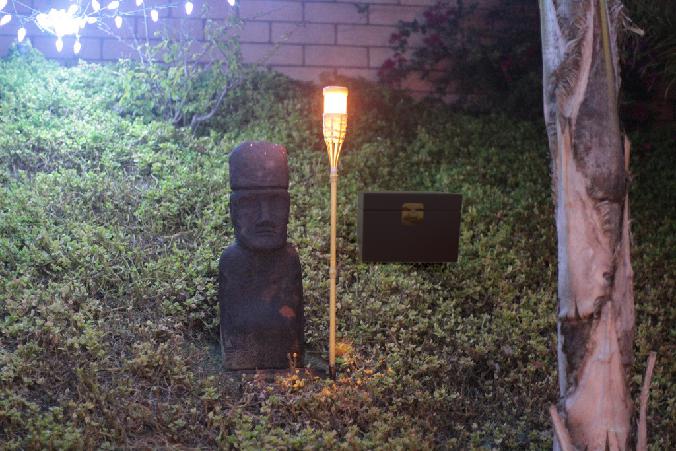}
\put(0,1){\fcolorbox{white}{white}{\includegraphics[width=0.33\linewidth,clip,trim=0 0 0 0]{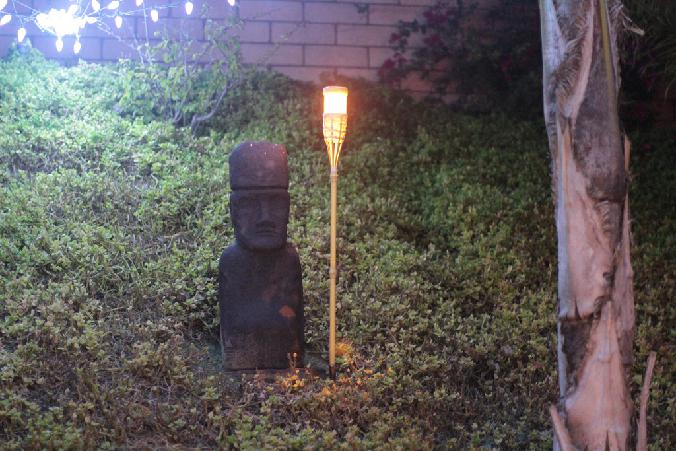}}}
\end{overpic}\\
\begin{overpic}[width=\linewidth,clip=true,trim=0 0 0 130]{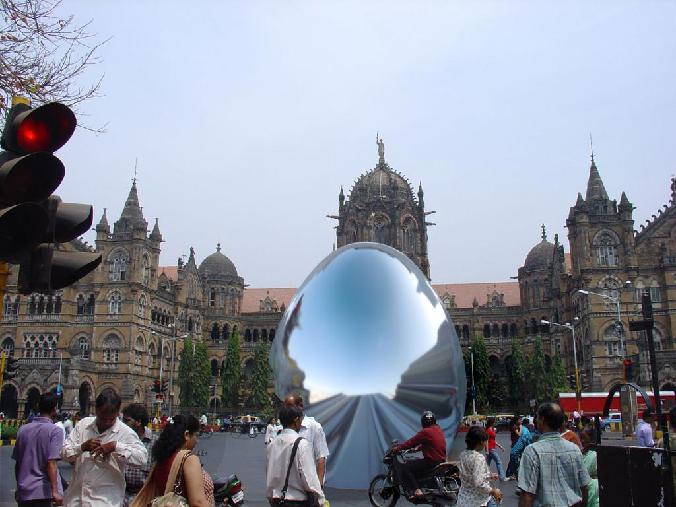}
\put(107,60){\fcolorbox{white}{white}{\includegraphics[width=0.33\linewidth,clip,trim=0 0 0 0]{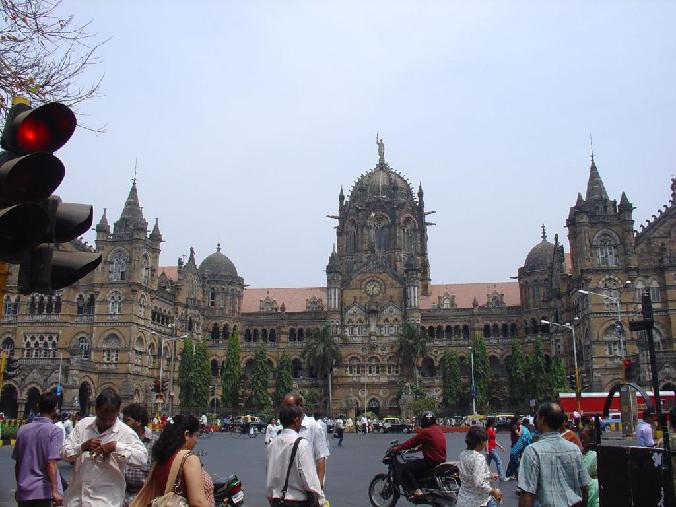}}}
\end{overpic}
\end{minipage}
\hfill
\begin{minipage}{.32\linewidth}
\begin{overpic}[width=\linewidth,clip=true,trim=0 70 0 0]{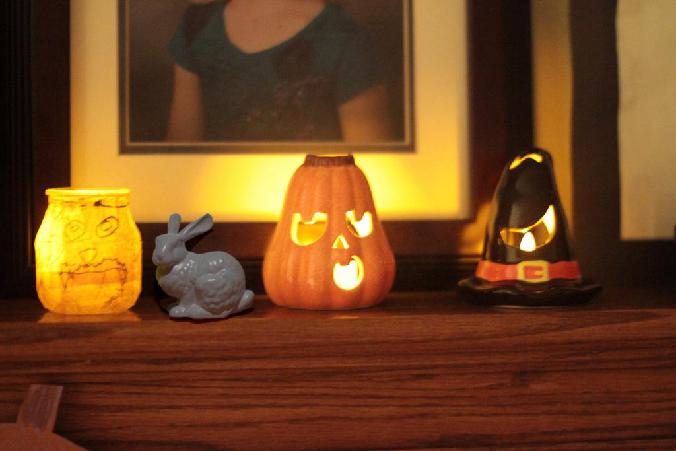}
\put(107,1){\fcolorbox{white}{white}{\includegraphics[width=0.33\linewidth,clip,trim=0 0 0 0]{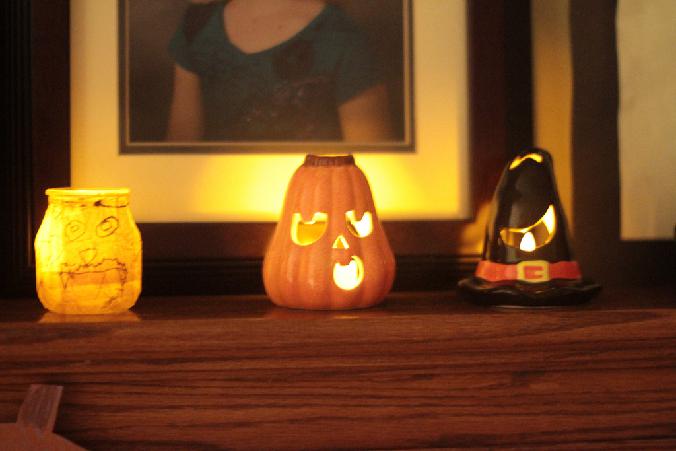}}}
\end{overpic}\\
\begin{overpic}[width=\linewidth,clip=true,trim=0 70 0 350]{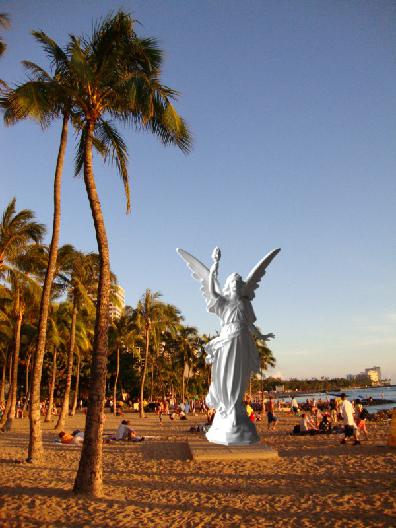}
\put(0,0){\fcolorbox{white}{white}{\includegraphics[width=0.33\linewidth,clip,trim=0 0 0 0]{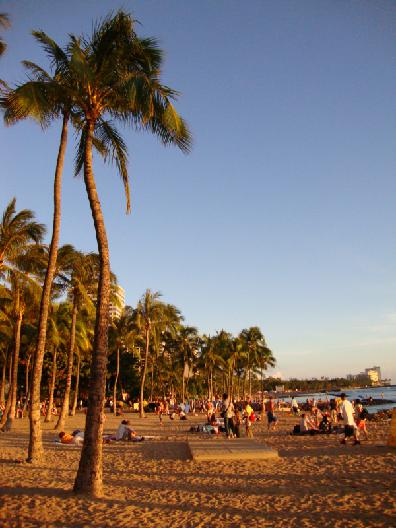}}}
\end{overpic}
\end{minipage}
\vspace{-1mm}
\caption{Failure cases. In some images, placement can be difficult due to inaccurate geometry estimates (top row), e.g. the dragon's shadow (top left) is not visible due to poor depth estimates, and the chest (top middle) rests on a lush, slanted surface, and our depth is flat in the region of the chest. Insertions that affect large portions of the image (e.g. statue; bottom left) or reflect much of the estimated scene (e.g. curved mirrors; bottom middle) can sometimes exacerbate errors in our scene estimates. Light estimates and color constancy issues lead to less realistic results (top right), especially outdoors (bottom right). Note that the people in the bottom right image have been manually composited over the insertion result. {\it Photo credits (left to right, top to bottom): {\footnotesize \textcopyright{}}Dennis Wong, {\footnotesize \textcopyright{}}Parker Knight, {\footnotesize \textcopyright{}}Parker Knight, {\footnotesize \textcopyright{}}Kenny Louie, Flickr user {\footnotesize \textcopyright{}}``amanderson2,'' and {\footnotesize \textcopyright{}}Jordan Emery.}
}
\label{fig:failure_examples}
\end{figure*}
%----------------------------------------------- Failure cases

%%%%%%%%%%%%%%%%%%%%%%%%%%%%%%%%%%%%%%%%%%%%%%%%
\begin{acks}
This work was funded in part by the NSF GRFP, NSF Award IIS 0916014, ONR MURI Award N00014-10-10934, and an NSF Expeditions Award, IIS-1029035. Any opinions, findings, and conclusions or recommendations expressed in this material are those of the author(s) and do not necessarily reflect the views of the National Science Foundation or of the Office of Naval Research.
\end{acks}

%%%%%%%%%%%%%%%%%%%%%%%%%%%%%%%%%%%%%%%%%%%%%%%%
{ %\small
\bibliographystyle{acmtog}
\bibliography{auto_illum}
}

\end{document}